\documentstyle[prc,epsfig,aps,preprint]{revtex}

\begin{document}

\date{\today}

\title{Collective response of nuclei : comparison between 
experiments and extended mean-field calculations}
\author{Denis Lacroix$^{a)}$, Sakir Ayik$^{b)}$ and Philippe Chomaz$^{c)}$}
\address{$^{a)}$ LPC/ISMRA, Blvd du Mar\'{e}chal Juin, 14050\ Caen, France  \\
$^{b)}$ {\it Tennessee Technological University, Cookeville TN38505, USA}\\
$^{c)}$ {\it G.A.N.I.L., B.P. 5027, F-14076 Caen Cedex 5, France } \\
\begin{abstract}
The giant monopole, dipole and quadrupole responses in 
$^{40}$Ca, $^{90}$Zr, $^{120}$Sn and $^{208}$Pb are investigated using 
linear response treatment based on a 
stochastic one-body transport theory. Effects of the coupling to low-lying
surface modes (coherent mechanism) and the incoherent mechanism due to
nucleon-nucleon collisions are included  beyond the usual mean-field description. 
We emphasize the importance of both mechanism in the fragmentation 
and damping of giant resonance. Calculated spectra are compared 
with experiment in terms of percentage of Energy-Weighted Sum-Rules in various 
energy regions. We obtained reasonable agreement in all cases.
A special attention as been given to the fragmentation of the Giant Quadrupole
Resonance in calcium and lead. In particular, the equal splitting of the 
$2^{+}$ in $^{40}$Ca is correctly reproduced. In addition, the appearance of
fine structure in the response $^{208}$Pb is partly  described by the calculations
in which the coherent mechanism play an important role.
\end{abstract}
}

\maketitle




\section{introduction}

Recent development of high resolution experiments
offers the possibility for a  deeper  
understanding of collective motion in quantum fermionic system
like nuclei. These experiments enable to determine the 
fragmentation of the nuclear response with a very high 
resolution up to few keV  \cite{Van91,Kuh81,Kam97-2,Lac00-2}.
Understanding of  the fine structure in the nuclear collective response, 
its fragmentation and damping mechanisms constitute a major challenge 
for theoretical models \cite{Ber83,Ber94,Dro90,Lac99}. One possible avenue is
the  
development of quantum transport theories for nuclear dynamics 
\cite{Goe82,Abe95}.

In dynamics of nuclear motion, one usually distinguishes
damping due to the coupling to the external and the internal degrees of freedom.
The former one gives rise to cooling of the system by evaporation of
particles, while the latter one leads to the dispersion of the 
well ordered motion through mixing with the internal degrees of freedom.
In the latter case, one can again distinguish
(i) the Landau damping due to spreading of the collective modes over 
non-collective particle-hole (p-h) excitation,  (ii) the coherent 
mechanism due to
coupling with low-lying surface modes \cite{Ber83,Bor81}, and 
(iii) the damping due to coupling with the incoherent 2p-2h states 
usually referred to as the collisional damping \cite{Ayi98,Yil99}.

Most investigations of the nuclear response carried out so far are 
based on either the coherent damping mechanism or the collisional 
damping.  The coherent mechanism is, particularly, important at low
temperature, and accounts for the main feature of fragmentation of 
the response \cite {Kam97-2,Bor81,Bor86,Gio98,Kam95,Kam97-1}. 
On the other hand, the collisional
damping is relatively weak at low temperature \cite{Deb92}, but its magnitude
becomes large with increasing temperature, as shown in recent 
calculations \cite{Lac98,Ayi00-1,Cho00,Lac00}.  In this work, we 
carry out investigations of nuclear collective response on the basis 
of  a  one-body  stochastic transport theory, which incorporates both 
the coherent mechanism and the collisional damping  in an consistent 
manner as demonstrated in \cite {Ayi00,AyiAbe} 
(also see \cite{Ayi94,Ivanov}).  We  calculate the giant 
monopole, dipole and quadrupole responses in $^{40}$Ca, $^{90}$Zr, 
$^{120}$Sn and $^{208}$Pb , and  compare the results  with experiment 
in terms of Energy-Weighted Sum-Rules distribution. We find that 
both mechanisms play important role for a proper description of
the fragmentation  and the damping of giant resonance excitations. 

In section II, we present a brief description of the linear response 
treatment of collective vibrations based  the stochastic one-body 
transport theory . In section III, we discuss  the details of the 
calculations and present the results and comparison with data in 
section IV. Finally, we give  the conclusions in section V.

\section{Linear Response Based on Stochastic Transport Theory}

\subsubsection{Stochastic transport equation}

In the stochastic transport theory, temporal evolution of the 
fluctuating single-particle density matrix ${\rho}(t)$
is determined by \cite{Abe95},

\begin{equation}
i\hbar \frac{\partial }{\partial t}{\rho}(t)-
[h({\rho}),{\rho}(t)]=K_I({\rho})+ \delta K(t) 
\label{ETDHF_full}
\end{equation}
where the left hand side corresponds to the mean-field evolution
in terms of the self-consistent mean-field Hamiltonian $h({\rho})$
expressed in terms of the fluctuating density, and the right hand side
arises from the correlations due to residual interactions. The first
term $K_I({\rho})$,which is usually referred to as the binary
collision term, describes the coupling of single-particle excitations
with more complicated two-particle two-hole states. It can be
expressed as, 

\begin{equation}
K_I({\rho} )=\int_{t_{0}}^{t}\left[ v,U_{12}(t,s)F_{12}(s)U_{12}^{\dagger}
(t,s)\right] ds  
\label{Eq:Ki}
\end{equation}
where $U_{12}(t,s)$ represents a product of the mean-field propagator
$U_{12}=U_{1}\otimes U_{2}$ with 
$\displaystyle U(t,s)=\exp \left( -i/\hbar
\int_{s}^{t}h(\rho (t^{\prime }))dt^{\prime }\right)$ 
and

\begin{eqnarray}
F_{12}=(1-{\rho}_{1})(1-{\rho}_{2}) 
v \widetilde{{\rho}_{1}{\rho}_{2}}-
\widetilde{{\rho}_{1}{\rho}_{2}}
v (1-{\rho}_{1})(1-{\rho}_{2}).
\end{eqnarray}
Here  $\widetilde{{\rho}_{1}{\rho}_{2}}$
Represents the anti-symmetric product of the single-particle
density matrices and $v$ denotes the residual
interactions. As seen from eq.(2), the collision term,
in general, involves memory effects due to the time
integration over the past history from an initial time $t_0$
to the present time $t$.  The second term in the right hand 
side of eq.(1) is the initial correlation term,

\begin{equation}
\delta K(\rho)=Tr_2 [v, \delta \sigma_{12}(t)]. 
\end{equation}
where $ \delta \sigma_{12}(t) =
U_{12}(t,t_0) \delta \sigma_{12}(t_0)U_{12}^{\dagger}(t,t_0)$
represents the propagation of the initial correlations from
$t_0$ to $t$. In the stochastic transport description, the 
initial correlations $\delta \sigma_{12}(t)$ are treated as
a Gaussian random quantity. Consequently, the initial correlation
term $\delta K(t)$ has a Gaussian distribution characterized by
a zero mean and a second moment, which can be determined in accordance
with the fluctuation-dissipation relation of the non-equilibrium
statistical mechanics. 

In the stochastic transport description, dynamical evolution is
characterized by constructing an ensemble of solutions of the
stochastic transport eq.(1). In this manner, the theory provides
a basis for describing the average evolution, as well as, dynamics
of density fluctuations. In the semi-classical approximation,
a number of applications have been carried out for description
of multi-fragmentation in heavy-ion collisions \cite{Gua96,Fra00}. 
Furthermore, as demonstrated in recent publications \cite {Ayi00,AyiAbe}, 
the stochastic
evolution involves a coherent dissipation mechanism arising from the
coupling of single-particle motion with the mean-field fluctuations.
This can
be shown by considering the average evolution of the single-particle
density matrix $ \rho(t)=\overline{ \rho}(t)$. The ensemble average
of eq.(1) is calculated by expressing the mean-field and the density
matrix as $h(\rho)=h(\overline{\rho})+ \delta {h}(t)$ and
${\rho(t)}=\overline{\rho}(t)+ \delta {\rho}(t)$, where  
$\delta{h}(t)$ and $\delta {\rho}(t)$ represent the 
fluctuating parts of the mean-field and the density matrix, 
respectively. Then, the ensemble averaging yields a transport
equation for the evolution of the average density matrix,

\begin{eqnarray}
i\frac{\partial}{\partial t}\overline{\rho}(t)-[h(\overline{\rho}),
\overline{\rho}(t)]
=K_I(\overline{\rho}) + K_C(\overline{\rho})
\label{etdhf_ci}
\end{eqnarray}
where $K_I(\overline{\rho})$ represents the incoherent collision term and
the additional term arises from the correlations of the mean-field
fluctuations and the density fluctuations,

\begin{equation}
K_C(\overline{\rho})=
\overline{\left[\delta {h}(t), \delta {\rho}(t) \right]}
\end{equation}
and it is referred to as the coherent collision term. For
small fluctuations around the average evolution, the density
fluctuations can be expressed in terms of RPA phonons, and 
the coherent term takes the form of a particle-phonon
collision term. As a result, 
eq.(5) provides an extended mean-field description, which goes 
beyond the extended Time-Dependent Hartree-Fock theory by including
a coherent collision term into the equation of motion in addition
to the incoherent one \footnote{In the following, we denote the average 
one-body density $\bar{\rho}$ by $\rho$.}.

\subsubsection{Linear response based of  extended mean-field theory}

In this section, we consider the small amplitude limit the transport
eq.(5) and give a brief description of the linear response
formalism including both the incoherent and the coherent damping terms.
A details description of the formalism can be found 
in recent publications \cite{Cho00,Ayi00}. 

The linear response of the system to an external perturbation
can be described by considering the small amplitude limit of the 
transport eq.(5). The small deviations of the density matrix 
$\delta \rho (t)=\rho (t)-\rho _{0}$ around a finite temperature
equilibrium state $\rho _{0}$ are determined by the linearized
form of the transport eq.(5), 
\begin{equation}
i\hbar \frac{\partial }{\partial t}\delta \rho -
[h_{0},\delta \rho ]-[\delta U+A,\rho _{0}]=
\delta K^I(\rho) + \delta K^C(\rho)  \label{Eq:d1}
\end{equation}
In this expression $A(t)=A\exp (-i\tilde{\omega}t)+{\rm h.c.}$ is a 
one-body harmonic excitation operator containing a small imaginary 
part $\tilde{\omega}=\omega +i\eta $, and  $\delta K_I(\rho)$ and
$\delta K_C(\rho)$  represents the linearized form of the non-Markovian 
incoherent and coherent collision terms, respectively. 

The steady state solution of eq. (\ref{Eq:d1}) can be obtained by using a
development in terms of the RPA functions, 
\begin{equation}
\delta \rho (t)=\left[ Q^{+},\rho _{0}\right] \exp (-i\tilde{%
\omega}t)+{\rm h.c.}
\end{equation}
where $Q^{+}=\sum_{\lambda >0}z_{\lambda }^{+}$ $%
O_{\lambda }^{\dagger }-z_{\lambda }^{-}O_{\lambda }$. 
In this expression, $ O_{\lambda }^{\dagger }$ and $O_{\lambda }$ are the creation and
annihilation operators of the RPA state $\lambda $ of energy $\hbar \omega
_{\lambda }$ , which are determined by the finite temperature RPA
equations \cite{Rin80},
\begin{eqnarray}
\hbar \omega_{\lambda }  O_{\lambda }^{\dagger }=[h_{0},Q _{\lambda
}^{\dagger }]+h_{\lambda }^{\dagger },
\label{RPA}
\end{eqnarray}
where $h_{\lambda }^{\dagger } = (\partial h / \partial \rho) \cdot  
\rho_{\lambda }^{\dagger }$.  Substituting the expression (8) into
eq.(7) gives rise to a set of coupled equations for the amplitudes
 $ z_{\lambda}^{+}$ and  $z_{\lambda }^{-} $ coefficients 
that can be recast into a matrix form \cite{Cho00},
\begin{equation}
\left( \hbar \tilde{\omega}-{\Huge \Sigma }\left( \tilde{\omega}\right) \right)
\left( 
\begin{array}{l}
z^{+} \\ 
z^{-}
\end{array}
\right) =\left( 
\begin{array}{r}
A \\ 
-A^{*}
\end{array}
\right)   \label{Eq:mat}
\end{equation}
where $z^{+}$ and $z^{-}$ are the amplitude vectors
with components $z_{\lambda}^{+}$ and $z_{\lambda}^{-}$, $A$ is the forcing
vector with components
$A_{\lambda }={\rm Tr}[O_{\lambda },A]\rho _{0}$ and 
$\Sigma \left(\tilde{\omega}\right)$ denotes the self-energy matrix. In the small
amplitude limit,  the self-energy can be separated into the incoherent part and the
coherent part, $\Sigma \left(\tilde{\omega}\right)= \Sigma_I \left(\tilde{\omega}\right) 
+ \Sigma_C \left(\tilde{\omega}\right) $.  
According to \cite{Cho00}, the expression of the incoherent part,  which also contains
the RPA energy,  is given by 
\begin{equation}
{\Huge \Sigma^I}_{\lambda\mu}\left( \tilde{\omega}\right) =\left( 
\begin{array}{lr}
\omega _{\lambda }\delta _{\lambda \mu }+{K}_{\lambda \mu }^{++}  
(\tilde{\omega})  & {K}_{\lambda \mu}^{+-} (\tilde{\omega})  \\ 
-{K}_{\lambda \mu }^{+-^{*}} (-\tilde{\omega}^{*})  & - \omega
_{\lambda }\delta _{\lambda \mu }-
{K}_{\lambda \mu }^{++^{*}} (-\tilde{\omega}^{*})  
\end{array}
\right)   \nonumber 
\end{equation}
In the Hartree-Fock representation, the elements of the incoherent self-energy
are given by,

\begin{equation}
{K}_{\lambda \mu }^{++}\left( \tilde{\omega}\right)  =-\frac{1}{4}
\sum_{ijk\ell }\frac{\left\langle k\ell \right| [O_{\lambda },v]\left|
ij\right\rangle \left\langle ij\right| [O_{\mu }^{\dagger },v]\left| k\ell
\right\rangle }{\hbar \tilde{\omega}-\Delta \varepsilon _{ijk\ell }}{\cal N}
_{ijk\ell }  
\end{equation}
and
\begin{equation}
{K}_{\lambda \mu }^{+-}\left( \tilde{\omega}\right)  =\frac{1}{4}\sum_{ijk\ell
}\frac{\left\langle k\ell \right| [O_{\lambda },v]\left| ij\right\rangle
\left\langle ij\right| [O_{\mu },v]\left| k\ell \right\rangle }{\hbar \tilde{%
\omega}-\Delta \varepsilon _{ijk\ell }}{\cal N}_{ijk\ell } 
\end{equation}
with
${\cal N}_{ijk\ell }=(1-n_{i})(1-n_{j})n_{k}n_{\ell }-n_{i}n_{j}
(1-n_{k})(1-n_{\ell })$ and  $\Delta \varepsilon _{ijk\ell }=\varepsilon
_{i}+\varepsilon _{j}-\varepsilon _{k}-\varepsilon _{\ell }$, where
$\varepsilon_{i}$ and $n_{i}$ denote energies and  Fermi-Dirac
occupation numbers of the single-particle states. The collisional self-energy is 
non diagonal, and therefore  it introduces a coupling between 
different RPA modes through their decay channels, so-called collisional 
coupling.  
In the following, we will neglect the non-diagonal part, which in general
introduces a small correction to the strength distributions.

According to \cite{Ayi00,AyiAbe}, the expression of the coherent 
self-energy is given by 

\begin{eqnarray}
\Sigma^C_{\mu}(\tilde{\omega})&=&\sum_{\lambda i j} 
\frac{ |<i|[Q_{\mu}, h_{\lambda}^{\dagger} ]|j>|^2 }
{ \hbar \tilde{\omega}-\hbar \omega_{\lambda}-\epsilon_j+\epsilon_i} 
{\cal M}_{\lambda,ij}  \\
& &-\sum_{\lambda i j} 
 \frac{ |<i|[Q_{\mu}, h_{\lambda} ]|j>|^2 }
{ \hbar \tilde{\omega} + \hbar \omega_{\lambda}-\epsilon_j+\epsilon_i }
{\cal M}_{\lambda,ji}
\label{Eq:coh}
\nonumber 
\end{eqnarray}
where
\begin{eqnarray}
{\cal M}_{\lambda,ij}=
(N_{\lambda}^0+1)(1-n_j^0)n_i^0-N_{\lambda}^0n_j^0 (1-n_i^0) 
\end{eqnarray}
and $N_{\lambda}^{0}$ denotes the finite temperature boson occupation 
factors for the RPA modes 
$N_{\lambda}^{0}=1/[exp( \hbar \omega_{\lambda}/T)-1]$. 
In general, the coherent self-energy is also non-diagonal, and
it couples different RPA modes. Here, we neglect this coupling and retain
only the diagonal part. The coherent mechanism, which arises from coupling of the 
single-particle excitations with the mean-field fluctuations in the stochastic
transport theory, correspond to the coherent mechanism described 
in \cite{Ber83,Bor81} and its finite temperature extension using the
Matsubara formalism  in  \cite{Bor86,Gio98}. 

We can deduce the response function associated with an excitation operator
$A$ , by calculating the expectation value  $<A>={\rm  Tr}A\delta \rho (t)$
with the help of the expression (8). The corresponding strength distribution
is obtained by the imaginary part of the response function and it can be
expressed as, 
\begin{equation}
S(\tilde{\omega})=-\frac{1}{\pi }Im\left( 
\begin{array}{ll}
A^{*}, & A
\end{array}
\right) \left( \hbar \tilde{\omega}-{\Huge \Sigma }\left( \tilde{\omega}\right)
\right) ^{-1}\left( 
\begin{array}{r}
A^{{}} \\ 
-A^{*}
\end{array}
\right) .
\end{equation}
The strength function includes both damping mechanisms,
i.e. the collisional damping due to coupling with the 
incoherent 2p-2h states and the coherent mechanism due to
coupling with a low-lying phonon and p-h states. 

In our previous studies, we investigated the nuclear collective
response in the basis of the incoherent damping mechanism. We  
found that at low temperature, in particular for light and
medium weight nuclei, the incoherent damping mechanism has a
sizeable influence on the strength functions, and it becomes more
important at higher temperatures. On the other hand, in particular
for heavy nuclei, the coherent mechanism due to coupling of 
giant resonance with phonons plus p-h states, plays a dominant
role for describing the properties of cold giant resonance.  
In this paper, we want to clarify the relative importance of the
incoherent and the coherent mechanisms in collective
response in cold nuclei. For this purpose, we present three different
calculations by incorporating only the coherent mechanism, only the
incoherent mechanism and including both mechanisms in to the 
calculations, and compare the results with the experimental data.

\section{Details of the calculation}

\subsection{RPA calculation}

In order to obtain the solution of equation (\ref{Eq:mat}), we first solve the
RPA equation (\ref{RPA}) in a discrete basis. In order to account partially
for the states in the continuum,  
particle and hole states are obtained by
diagonalizing the Hartree-Fock Hamiltonian in a large harmonic oscillator
representation \cite{Van81} which includes respectively 12 major shell 
for $^{40}$Ca and 15 major shells for other nuclei. 
We use a fixed imaginary part for the forcing frequency 
$\eta = \eta_s = 0.5$ MeV. The Hartree-Fock and RPA calculations
are performed using the effective Skyrme force SLy4\cite{Cha98}.
We use the standard excitation operators  for isoscalar and 
isovector resonances for $L \ge 1$ (for a review see \cite{Kam97-1}),
\begin{eqnarray} 
\begin{array} {l}
A_{LM} = \frac{Z}{A}\sum_{i=1}^{A} r_i^L Y_{LM} \\
\\
A_{LM} = \frac{N}{A}\sum_{i=1}^{Z} r_i^L Y_{LM} -
 \frac{Z}{A}\sum_{i=1}^{N} r_i^L Y_{LM}
\end{array}
\end{eqnarray}
where $Y_{LM}$ are the spherical harmonics, and for isoscalar
giant monopole resonance, we employ
\begin{eqnarray} 
A_{00} = \frac{Z}{A}\sum_{i=1}^{A} r_i^2 Y_{00} 
\end{eqnarray}
The energy-weighted sum rule 
(EWSR) is given by,
\begin{eqnarray}
m_1 = \sum_\lambda \hbar \omega_\lambda 
\left| \left<0\left|A\right| \lambda \right>\right|^2 
\label{sum_sum}
\end{eqnarray}
When the states $\lambda$ are specified in the RPA, it
can be calculated from the Hartree-Fock
ground state according to
$ m_1=\frac 12<[F^{\dagger },[H,F]]>_0$. 
For the Skyrme interactions, it leads to the 
following expression, 

\begin{eqnarray} 
\left\{
\begin{array} {l}
m^{GMR}_1 = \frac{2\hbar^2}{m} \frac{Z^2}{A} \left< r^2 \right>_{HF} \\
\\
m^{GDR}_1 = \frac{9}{4 \pi} \frac{\hbar^2} {2m} \frac{NZ}{A} (1-\kappa ) \\
\\
m^{GQR}_1 = \frac{50}{4\pi } \frac{\hbar ^2}{2m}\frac{Z^2} A
\left\langle
r^{2}\right\rangle _{HF}
\end{array}
\right.
\label{sum_123}
\end{eqnarray}
where $\left\langle r^{2}\right\rangle _{HF}$ denotes the root-mean 
square radius (rms) obtained from the Hartree-Fock ground state. 
In the case of the Giant
Dipole Resonance, the Thomas-Reich Kuhn (TKR) sum rule is violated 
due to the non-local term in Skyrme forces, and the modification
factor is giving by \cite{Sag84,Bar85},
\begin{equation}
\kappa =\frac{2m}{\hbar ^2}\left[ t_1(1+\frac 12x_1)+t_2(1+\frac
12x_2)\right]  \nonumber \\
\frac{1}{A} \int \rho_n(r) \rho_p(r) d^3 r, 
\end{equation}
where $\rho_n$ and $\rho_p$ are the neutron and proton one-body density.

In the following, we compare the result of calculations
with the experimental EWSR by employing the standard 
expression and parameters of the sum-rule \cite{Sat87,Hor95}. 
In the standard approach, the rms radius is approximated 
using a Wood-Saxon shape 
for the one-body density, which leads to the following 
expression,
\begin{eqnarray}
\left< r^2 \right>_{WS} = \frac{3}{5} R_0 \left( 1.+ \frac{7}{3}
\left[\frac{\pi a}{c}\right]^2 \right)
\label{rws}
\end{eqnarray}
where $R_0$ correspond to the surface position and 
a is the diffuseness.
Different parameters used in the calculations are reported 
in table \ref{tab:rms}. In table \ref{tab:sum}, we compare the sum rules
obtained by the parameterization  (\ref{rws}) and $\kappa=0$ and those
results obtained  from  the RPA calculations of (\ref{sum_sum}).
The smallness of the difference insures the quality of the RPA calculations.

\begin{table}[tbp]
\begin{center}
\begin{tabular}{cccccc}
Nucleus     & $R_0$  & a  & 
$\left< r^2 \right>_{WS}^{1/2}$  &
$\left< r^2 \right>_{HF}^{1/2}$ & $\kappa$ \\
&&&&& \\
& (fm) & (fm) & (fm) & (fm) &  \\
\hline
$^{40}$Ca \protect{\cite{You97}}  & 3.65   & 0.55  & 3.49 & 3.40 & 0.160 \\  
$^{90}$Zr \protect{\cite{Hor95}}  & 4.90    & 0.515 & 4.25 & 4.26 & 0.177 \\  
$^{120}$Sn \protect{\cite{Hor95}} & 5.55   & 0.515 & 4.71 & 4.70 & 0.176 \\
$^{120}$Pb \protect{\cite{Hor95}} & 6.67   & 0.545 & 5.55 & 5.55 & 0.180 \\
\end{tabular}
\end{center}
\caption{Density shape parameters used in the calculations of the
sum-rules. Hartree-Fock rms obtained with the Sly4 force are very 
close to those obtained with the Wood-Saxon parameterization. }
\label{tab:rms}
\end{table}

\begin{table}[tbp]
\begin{center}
\begin{tabular}{crrr}
Nucleus & 0$^{+}$ (MeV. fm$^4$) & 1$^{-}$ (MeV. fm$^2$)& 2$^{+}$ 
(MeV. fm$^4$)  \\
\hline 
$^{40}$Ca  & 10103 (9290)  & 148 (144) & 10050 (9226) \\ 
& & & \\ 
$^{90}$Zr   & 26664 (26025)  & 330 (325) & 26523 (22556) \\
& & & \\ 
$^{120}$Sn  & 38295 (37026) &  433 (427) &  38092 (26477) \\
& & & \\ 
$^{208}$Pb  &  82633 (84572) & 738 (748) &  82197 (76034) \\
\end{tabular}
\end{center}
\caption{$m_1$ sum rule obtained from standard parameterization 
of the rms radii. 
Sum rule calculated from RPA are reported in parenthesis.}
\label{tab:sum}
\end{table}
 
\subsection{Computation of self-energies}

\subsubsection{Coherent mechanism.}
 
In order to incorporate the effect of coupling to surface modes, 
we calculate the RPA response for multipolarities up to $L=5$. 
The coherent self-energy  given by eq. (14) is then 
calculated by coupling collective 
states to low-lying states that exhaust at least 1$\%$ 
fraction of the EWSR. Energies and EWSR of 
collective modes used in the calculation are 
reported in table \ref{tab:coll}. We note that, 
in particular, energies of collective
$3^{-}$ states are overestimated in the RPA calculations. The
percentage of the EWSRs are normalized to the RPA sum-rules.
In the calculations of the
matrix elements in eq.(14), we employ the full SLy4 
interaction.
\begin{table}[tbp]
\begin{center}
\begin{tabular}{cccc}
\hline \\
Nucleus & J$^{\pi}$ & E (MeV) &  \% EWSR \\
\hline \\
$^{40}$Ca & 0$^{+}$ & 17.6  & 10.4 \\
          & 0$^{+}$ & 19.1  & 12.5 \\
          & 0$^{+}$ & 20.7  & 18.1 \\
          & 0$^{+}$ & 22.0  & 18.0 \\
          & 0$^{+}$ & 24.4  & 15.5 \\
          & 1$^{-}$ & 16.7  & 19.5 \\
          & 1$^{-}$ & 17.8  & 15.0 \\
          & 1$^{-}$ & 18.6  & 22.8 \\
          & 2$^{+}$ & 17.1  & 74.9 \\
          & 3$^{-}$ & 5.3   & 9.5 \\
          & 3$^{-}$ & 7.3   &  10.8 \\
          & 5$^{-}$ & 5.2   &  3.7 \\
          &&& \\
\hline
          &&& \\
$^{90}$Zr & 0$^{+}$ & 17.8  & 35.1\\
          & 0$^{+}$ & 18.6  & 17.1 \\
          & 1$^{-}$ & 15.7  & 50.1 \\
          & 1$^{-}$ & 17.9  & 14.9 \\
          & 2$^{+}$ & 5.4  &  5.4 \\
          & 2$^{+}$ & 16.3  & 13.5 \\
          & 2$^{+}$ & 16.4  & 54.4 \\
          & 3$^{-}$ & 3.7  &  4.9 \\
          & 3$^{-}$ & 7.9  &  19.8 \\
          & 3$^{-}$ & 9.4  &  8.9 \\
          &&& \\
\hline
          &&& \\
$^{120}$Sn & 0$^{+}$ & 16.8  & 30.7 \\
          & 0$^{+}$ & 18.4  & 42.4 \\
          & 1$^{-}$ & 14.1  & 20.1 \\
          & 1$^{-}$ & 14.5  & 13.7 \\
          & 1$^{-}$ & 16.9  & 15.7 \\
          & 2$^{+}$ & 6.0  &  3.5 \\
          & 2$^{+}$ & 16.5  & 23.0 \\
          & 2$^{+}$ & 16.6  & 24.0 \\
          & 3$^{-}$ & 3.8  & 6.7 \\
          & 3$^{-}$ & 6.9  & 7.2 \\
          & 3$^{-}$ & 7.6  & 18.0 \\
          & 5$^{-}$ & 7.0  & 2.7 \\
          & 5$^{-}$ & 8.3  & 1.1 \\
          & 5$^{-}$ & 9.5  & 5.8 \\
          &&& \\
\hline
          &&& \\
$^{208}$Pb & 0$^{+}$ & 14.4  & 33.5 \\
           & 0$^{+}$ & 14.7  & 18.8 \\
           & 0$^{+}$ & 15.0  & 19.5 \\
           & 0$^{+}$ & 16.4  & 14.3 \\
           & 1$^{-}$ & 12.8  & 13.7 \\
           & 1$^{-}$ & 13.2  & 21.2 \\
           & 1$^{-}$ & 13.7  & 13.8 \\
           & 2$^{+}$ & 3.7  &  1.4 \\
           & 2$^{+}$ & 6.1  &  10.2\\
           & 2$^{+}$ & 13.0  & 67.5 \\
           & 3$^{-}$ & 4.9  &  21.9 \\
           & 3$^{-}$ & 6.7  &  2.0\\
           & 3$^{-}$ & 7.2  &  5.7 \\
           & 3$^{-}$ & 9.1  &  1.3 \\
           & 4$^{+}$ & 6.6  &  4.0 \\
           & 4$^{+}$ & 8.9  &  3.1 \\
           & 4$^{+}$ & 9.3  &  3.1 \\
           & 4$^{+}$ & 9.6  &  3.1 \\
           & 5$^{-}$ & 4.8  &  1.3 \\
           & 5$^{-}$ & 5.7  &  2.3 \\
           & 5$^{-}$ & 5.9  &  1.1 \\
           & 5$^{-}$ & 6.7  &  1.1 \\
           & 5$^{-}$ & 7.7  &  2.5 \\
           & 5$^{-}$ & 8.2  &  1.7 \\
           & 5$^{-}$ & 9.3  &  1.6 \\
           & 5$^{-}$ & 9.5  & 1.9 \\
\end{tabular}
\end{center}
\caption{Collective modes and associated EWSR 
obtained from RPA that are included in the calculations.}
\label{tab:coll}
\end{table}

\subsubsection{Incoherent mechanism}

We have shown in previous calculation that Skyrme interaction 
are not adequate to compute the incoherent part of the self-energy 
due to the presence of high momentum component
(also, see \cite{Yil99}). As in previous application 
\cite{Lac98},
we use a modified Skyrme interaction which is obtained by 
introducing a Gaussian cut-off factor in the matrix elements 
of the Skyrme force $v_{S}$,  
\begin{equation}
<ij|v|kl>=<ij|v_{S}|kl>\cdot \exp \left( -\frac{\beta^2|<{\bf q}^{2}>|}{\hbar
^{2}}\right)   \label{range}
\end{equation}
in this expression, $\beta$ describes an effective range of the interaction
and the quantity $<{\bf q}^{2}>$
provides a measure for the relative momentum which is defined by the
relation $<ij|\delta ({\bf r})|kl><{\bf q}^{2}>=<ij|{\bf q}^{2}\delta ({\bf r%
})+\delta ({\bf r}){\bf q}^{2}|kl>$ with ${\bf r}={\bf r}_{1}-{\bf r}_{2}$
and ${\bf q}=({\bf p}_{1}-{\bf p}_{2})/2$. A quantitative discussion of the
influence of $\beta$ can be found in \cite{Lac00}. In the following, we use
$\beta=1.4$ fm. The size of the HO basis used to expend the particle and hole (HF) states
have been chosen large enough to ensure a convergency of the results with
in few per cent. 

\section{Results}

\begin{figure}[tbph]
\begin{center}
\epsfig{file=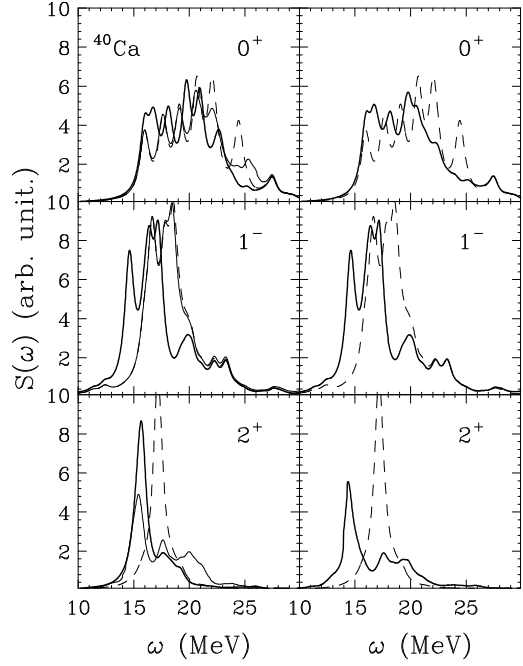,width=7.cm}
\end{center}
\caption{Calculated strength distributions for GMR (top),
GDR  (middle) and  GQR (bottom) in $^{40}$Ca.
Left: strengths obtained in  the RPA (dashed lines), with the coherent 
mechanism (thin lines) and the incoherent mechanism (thick lines).  
Right: comparison between the RPA (dashed lines) and the extended 
RPA (thick lines), which includes both the coherent and the incoherent 
damping mechanisms.}
\label{fig:ca1}
\end{figure}

\begin{figure}[tbph]
\begin{center}
\epsfig{file=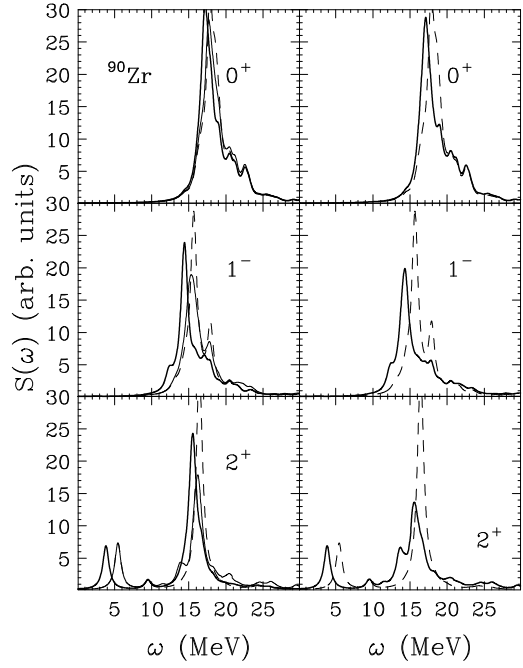,width=7.cm}
\end{center}
\caption{Same as figure \protect{\ref{fig:ca1}} for $^{90}$Zr. }
\label{fig:zr1}
\end{figure}

\begin{figure}[tbph]
\begin{center}
\epsfig{file=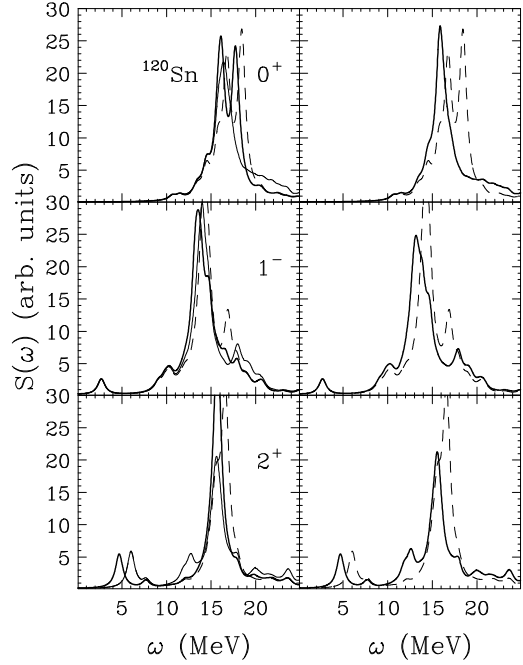,width=7.cm}
\end{center}
\caption{Same as figure \protect{\ref{fig:ca1}} for $^{120}$Sn. }
\label{fig:sn1}
\end{figure}

\begin{figure}[tbph]
\begin{center}
\epsfig{file=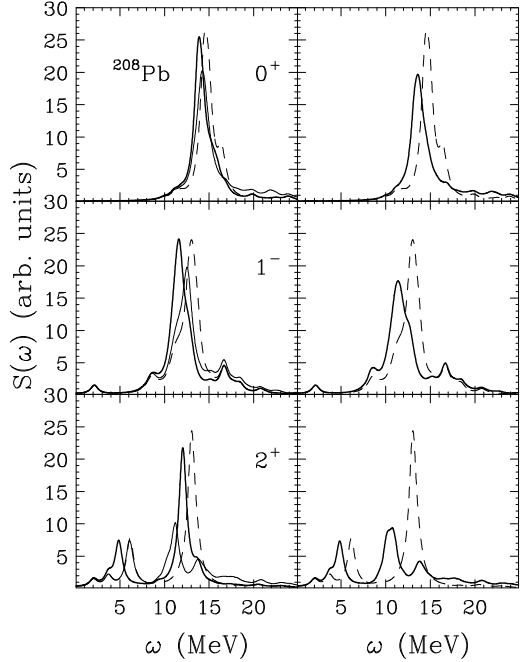,width=7.cm}
\end{center}
\caption{Same as figure \protect{\ref{fig:ca1}} for $^{208}$Pb. }
\label{fig:pb1}
\end{figure}

We carry out calculations of  strength functions for the 
giant monopole (GMR), 
the giant dipole (GDR) and the giant quadrupole (GQR)
response at zero temperature. The resulting strength 
distributions are presented for $^{40}$Ca (figure \ref{fig:ca1}), 
$^{90}$Zr (figure \ref{fig:zr1}), $^{120}$Sn (figure \ref{fig:sn1}) 
and $^{208}$Pb (figure \ref{fig:pb1}). 
The left panels of figures show the result of  RPA calculations  (dashed 
lines) and the calculations performed by including only the coherent self-energy
(thin lines) and only the  incoherent self-energy (thick lines). The result of
calculations performed by including both  self-energies are shown on the right
panel of  figures (thick lines).

\subsection{Interplay between incoherent and coherent mechanism}

In order to quantitatively discuss the effects of different 
contributions, it is useful to compute moments 
of the strength in a given energy interval,
\begin{eqnarray}
m_{i-[E_{min},E_{max}]} = \int^{E_{max}}_{E_{min}} \left(\hbar \omega\right)^i
S(w)
d \omega 
\end{eqnarray}
From these moments, we can define various mean energies 
$\overline{E_i} = m_i / m_{i-2}$.
An estimation of the spreading of the strength is given 
by the width 
$\displaystyle \overline{\Gamma} = \sqrt{ m_2/m_0 - \left( m_1/m_0 \right)^2}$.
In figure \ref{fig:de}, the difference $\Delta E  = \overline{E_1} - 
\overline{E_1}^{RPA} $ between the mean energy 
obtained in different calculations and the mean-energy
calculated in the RPA  is  plotted as a function of the mass number 
for the different multipolarities. In figure, calculations by including
the coherent mechanism, the incoherent mechanism and the coherent plus
incoherent are indicated by dashed lines, dashed-dotted lines and
solid lines,  respectively.
In figure \ref{fig:dg}, a similar plot is presented for the deviation of the width 
from the RPA response,  $\Delta \Gamma = \overline{\Gamma} - 
\overline{\Gamma}^{RPA} $.
In these calculations, moments of the strength functions are 
evaluated over  the energy interval 0-40 MeV. From the results
of calculations, we can draw the following conclusions:

\begin{figure}[tbph]
\begin{center}
\epsfig{file=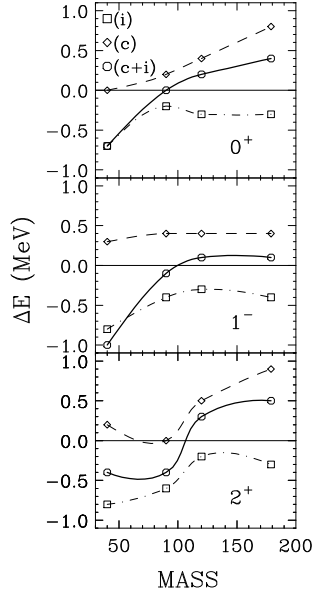,height=8.cm}
\end{center}
\caption{Variation of the mean-energy $\Delta \overline{E} = \overline{E_1} -
\overline{E_1}^{RPA}$ calculated  in the energy interval  0-40 MeV for 
GMR (top),  GDR (middle) and GQR (bottom) as a function of mass number. 
Calculations performed by including the coherent mechanism, the 
incoherent mechanism and both coherent and incoherent mechanisms
are indicated by dashed lines, dash-dotted lines and solid lines, respectively. }
\label{fig:de}
\end{figure}

\begin{figure}[tbph]
\begin{center}
\epsfig{file=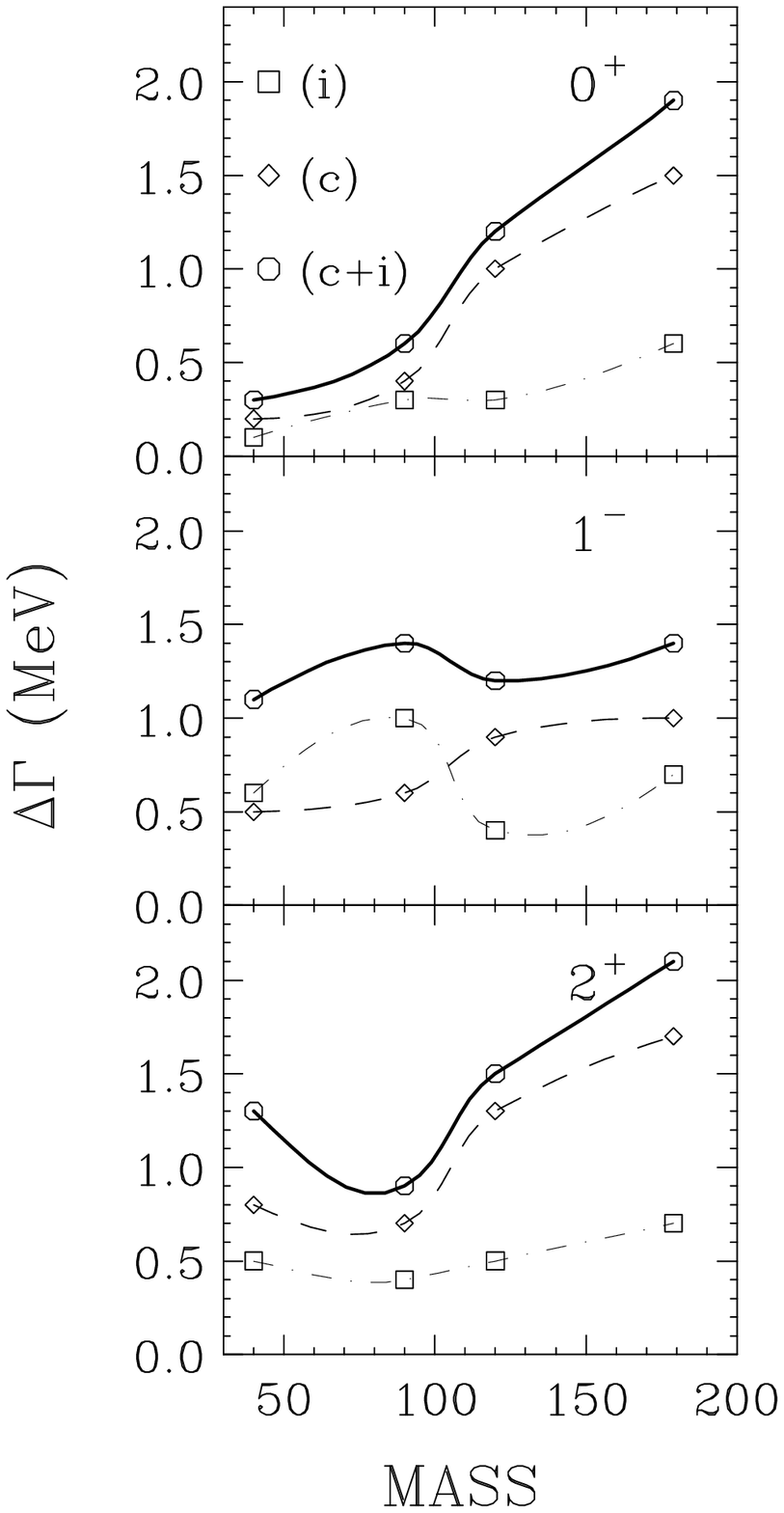,height=8.cm}
\end{center}
\caption{Variation of the mean-width 
$\Delta \overline{E} = \overline{E_1}-\overline{E_1}^{RPA}$ 
calculated  in the energy interval  0-40 MeV for GMR (top),  
GDR (middle) and GQR (bottom) as a function of mass number. 
Calculations performed by including the coherent mechanism, the 
incoherent mechanism and both coherent and incoherent mechanisms
are indicated by dashed lines, dash-dotted lines and solid lines, 
respectively.} 
\label{fig:dg}
\end{figure}

\begin{figure}[tbph]
\begin{center}
\epsfig{file=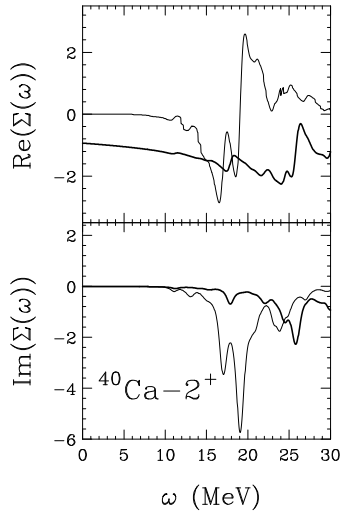,height=7.cm}
\end{center}
\caption{The real (top) and the imaginary (bottom) part of the 
coherent (thin line) and the incoherent (thick line) part of 
the self-energies for the GQR in $^{40}$Ca.}
\label{fig:self}
\end{figure}

\begin{itemize}
\item {\bf Shift of mean-energy: }
\begin{itemize}
\item The incoherent mechanism induces a reduction of the 
mean energy, while the coherent part acts in the opposite way.
The origin of this phenomenon can be found by 
looking carefully in figures from \ref{fig:ca1} to \ref{fig:pb1}. 
Indeed, we note that both coherent and incoherent self-energies
induces a shift of the main peaks towards lower energy. However, 
at the same time,  a part of strength is shifted toward higher energy 
in the coherent case, which gives rise to a 
global increase of the average mean-energy.
Such a behavior can be understood by looking at self-energies
themselves. An example, in figure \ref{fig:self}, the coherent 
(thin lines) and the incoherent (thick lines) 
self-energies are shown  for the GQR in $^{40}$Ca. We see that 
the energy dependence of the real  part of the self-energy is different in
two different mechanisms. While the incoherent mechanism induces
a global shift of the strength towards lower energies, the real part of the 
coherent self-energy change of sign in the vicinity of the collective energy.
This introduces  a shift toward lower energy of the low energy 
part of the strength while the high energy part is pushed towards higher
energies. In some cases, we may even expect that a single resonance is
splitted into two peaks, as it happens  for the GQR in $^{40}$Ca.

\item For the GQR and GMR in lighter nuclei, the shift introduced by the
incoherent mechanism  is stronger that the coherent one. On contrary, 
the tendency goes the opposite direction for heavier nuclei. 
For the GDR , both effect are comparable.  

\item In all cases, the effect of the incoherent mechanism is of the same order of
magnitude as that of  the coherent mechanism and can not be neglected 
in contrast to the usual assumption \cite{Deb92}.

\end{itemize}
\item {\bf Increase of spreading width: }
\begin{itemize}

\item Both the coherent and incoherent self-energy induces an increase
of the spreading.

\item In the case of the GQR and GMR, the coherent damping 
is always much larger than the incoherent one while for the 
GDR both are of same 
order. For instance, in the GQR case, where the strength
is in general not Landau fragmented, we can see that the coupling to surface 
modes induces a splitting of the main pic of the RPA into different peaks.
In the calcium case, this effect can be related to 
to the presence of two collective low-lying 3$^{-}$ states strongly coupled 
to the GQR.

\item The fact that the coherent mechanism induces a larger damping can be
seen by looking at bottom panel  of figure \ref{fig:self}.
In this particular example, we see that the imaginary part of the self-energy
is larger for the coherent mechanism than for the incoherent case, which gives 
a larger damping width.

\item When the strength is already largely Landau fragmented (like in the
GMR of $^{40}$Ca) both incoherent and coherent effects seems almost 
negligible.

\item  The magnitude of the coherent mechanism becomes larger 
for heavier nuclei.

\end{itemize}
\item {\bf Additivity of coherent and incoherent effects: }

\begin{itemize}
\item The shift in the mean energy, $\Delta E$, and the increase of the
width, $\Delta \Gamma$, are approximately 
given by the sum of  these quantities obtained by considering
the coherent and incoherent mechanisms separately. 
\end{itemize}
\end{itemize} 
 
The effects of both, coherent as well as incoherent mechanisms appears 
to be more significant
for the lighther nuclei as seen from figures 1-4. This may be due to
the fact that in ligther nuclei a large fraction of nucleons
resides in the vicinity of nuclear surface relative to the heavier ones,
where the predominant effects of both coherent as well as incoherent
mechanisms occur. Moreover, in light nuclei the energy of the 
collective states is higher leading to an increase of the 
damping width.

\subsection{Fragmentation of response}

With the high precision experiments, it  is possible to determine fragmentation 
and fine structure of the strength functions. 
In order to characterize systematically the shape of strength functions, we
present  properties of giant resonances spectra in tables 
\ref{tab0p:Ca40} - \ref{tab2p:Pb208}. Depending on the fragmentation of
each response, we report average mean energies and width for 
different energy intervals. When the strength is divided into several main
peaks, we  consider energy intervals around the main 
peaks. Besides the average properties, we also report the main peak 
positions.  We emphasize that, in particular for the GQR response, the
coherent mechanism induces an additional fragmentation 
of the strength. The incoherent mechanism also introduces such a 
fragmentation, but it is much weaker than the coherent effect. 
The incoherent damping strongly the peak positions.
In any case, for a proper description of the fragmentation and the
fine structure of the strength distributions, both the coherent and 
the incoherent mechanisms should be taken into the description.

\subsection{Low-lying states}

The RPA
calculations, most often, overestimate
the mean-energy of low-lying states. We find 
that the incoherent mechanism reduces systematically the mean-energy 
of the GQR states for medium and heavy nuclei. In table \ref{tab2p:Zr90},
\ref{tab2p:Sn120} and \ref{tab2p:Pb208}, we can see that the mean-energy of the 
low-lying 2$^{+}$ states is shifted by -1.3 MeV for $^{90}$Zr and $^{120}$Sn,
and by -0.7 MeV in $^{208}$Pb.    
Such a shift is absent in the calculations with the coherent damping mechanism, 
while it remains when the both mechanisms are included into the description.

\subsection{Comparison with experiment}

When the strength is highly fragmented, a direct comparison to 
experimental data is hardly possible. 
Experiments are often analyzed
using a fitting plus folding procedure  of spectra which mix different 
multipolarities\cite{Sat87}. From this procedure, one extract 
energy ($E_\alpha$) , width ($\Gamma_\alpha$) 
and percentage of  the EWSR ($\left(\%EWSR\right)_\alpha$). 
In order to compare with experimental data, we convert the experimental
data into  percentages of  the EWSR in given energy intervals, 
which is determined according to

\begin{eqnarray}
\begin{array} {ll}
\left(\% EWSR \right)_{\left[E_{min},E_{max}\right]} 
= & \sum_\alpha \footnotesize{\left(\%EWSR\right)_\alpha}   \\ 
& \\
& \times \int_{E_{min}}^{E_{max}} 
  \frac{\Gamma_\alpha/{2\pi}}{\left(E-E_\alpha\right)^2 - \Gamma^2_\alpha/4}
\end{array}
\end{eqnarray}  
In this expression, $\alpha$ runs over
different states of considered multipolarity, and a 
Lorentzian shape is assumed for each mode in the
calculations. We note that  Gaussian shapes rather than
Lorentzian only slightly  change the reported values.  
In tables \ref{tab0p:Ca40} - \ref{tab2p:Pb208}, we compare 
the results of our calculations for the percentage of  EWSR with experiments 
in the energy interval around the peak energy of the corresponding
giant resonance. In all cases,  our calculations provide a good description
for the experiments.  In general, the introduction of
coherent and incoherent mechanisms gives a better description of available 
experimental data. However, in some cases, the percentage of the EWSR obtained 
in RPA already gives the good order of magnitude. We pay particular attention
to the GQR excitations in $^{40}$Ca and $^{208}$Pb, 
since both have been extensively 
studied experimentally and have given long-standing discussions \cite{Van91}.

\subsubsection{Splitting of the $2^{+}$ resonance in $^{40}Ca$}

The GQR response in $^{40}$Ca is known to be split into two components 
with energy around 13.5 and 18 MeV with almost an equal fraction of the 
EWSR (around 30 $\%$ to 40 $\%$ for both peaks). The description of this
fragmentation by microscopic calculations is a problem.
Only recently \cite{Kam95,Kam97-1}, microscopic calculations assuming  
ground state correlations and coupling to low lying states
reproduce a global splitting. However, these calculations describe  the global 
trend of the response and do not provide an explanation for the equal partition of
the strength.

Looking at table \ref{tab2p:Ca40}, we see that, our calculation with the 
coherent and incoherent mechanism not only
reproduce splitting of the strength into two main components (see figure
\ref{fig:ca1}) but also give rise to an equal splitting around the main peak
(31 $\%$ in the interval 10-16 MeV and 33.6 $\%$ in the interval 
16-22 MeV),  that matches with the experimental data.
When only the coherent self-energy is included, the calculations 
can not reproduce the splitting, but give the percentage of the 
EWSR, which are comparable to those obtained 
in ref. \cite{Kam97-1} i.e. a too high percentage of the  EWSR for 
the second peak and a too low for the first one. 

This particular example demonstrates the necessity of taking both coherent 
and incoherent damping mechanisms at the same time and illustrates the
complementarity of the two effects. Indeed, without the 
coherent mechanism splitting of the strength is not 
found while without the incoherent damping the EWSR is not reproduced.
 
\subsubsection{Fine structure in $^{208}$Pb}

The calculated strength obtained in our microscopic calculations
for the GQR of $^{208}$Pb is displayed in bottom panel of figure 
\ref{fig:pb1}. As in the case of $^{40}$Ca,
global shape of the strength exhibits a splitting of the GQR response 
into two main peaks at 10.8 MeV and 13.8 MeV. The first peak is well known
and is correctly reproduced by our calculations, while the RPA alone
does not give a correct description.  However the fraction of the EWSR
is slightly  smaller as compared to the experiment 
(see bottom of table \ref{tab2p:Pb208}).
The second peak has never been observed but is also present 
in second RPA calculations \cite{Lac00-2,Dro90}. 

\begin{figure}[tbph]
\begin{center}
\epsfig{file=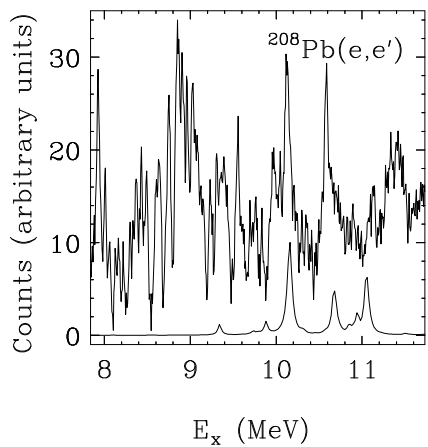,height=6.cm}
\end{center}
\caption{Thin line: the strength function for GQR in $^{208}$Pb,
which is calculated including the coherent and the incoherent 
self-energies with a smoothing parameter $\eta = \eta_s = 0.025$ MeV.  
Thick line: the experimental spectra obtained in the inelastic 
electron scattering experiment. }
\label{fig:pbfine}
\end{figure}

Our calculation, which assume a rather large value of the smoothing parameter
only gives a global shape of the strength distributions. In order to reveal the fine
structure on top of the global shape, we  also perform calculation s
with smaller smoothing parameter $\eta = \eta_s = 0.025$ MeV which corresponds
to experimental resolution. The corresponding strength distribution is 
presented in figure \ref{fig:pbfine} for the collective
energy region  7.6 - 11.8 MeV. The calculated response is compared with
the inelastic electron scattering data  \cite{Kuh81}.  
This experimental data presents a well-defined fine structure 
which is also observed with a one-to-one correspondence 
in (p,p') experiments \cite{Lis91,Kam97-2}. In figure \ref{fig:pbfine},
we see that the calculations agree with the part of  spectral properties 
of the peaks in the vicinity of the collective energy. However, we note that below 9 MeV, 
the fine structure  is almost absent in our description. The peak positions observed 
experimentally and obtained in our calculations are reported in table \ref{tab:fine}.
We can see from this table that fine structures are already present 
in the coherent case while they are absent in the incoherent one. 
When both effects are included, it seems that part of the peaks are 
perfectly located as compared to recent (p,p') experiments.
It has been recently discussed that other peaks might be coming from  
dipole excitations \cite{Dro00}. It is also possible that missing 
peaks might be due to the fact that
part of the two-body correlations are neglected in the present 
description or coming from higher order correlations.

\section{Conclusion}

In this article, we carry out a systematic investigation the effect of coherent 
and incoherent damping mechanisms on the collective response in spherical 
nuclei at zero temperature. Our calculations indicate that both mechanisms
play important roles in a proper description of   the nuclear collective response.
An extensive comparison with experimental data is presented 
in terms of  the fraction of exhausted EWSR for the GMR, GQR and
GDR for a number of  nuclei. We show that the presented 
calculations are in reasonable agreement with the observed 
collective response.
A special attention has been given to the GQR response in calcium and lead 
nuclei where a large amount of experimental and theoretical work exists.
In particular, we show that, while the usual mean-field theory is unable 
to explain the equal splitting of the 2$^{+}$ state, the inclusion of both 
coherent and incoherent damping mechanism provides an explanation 
for  fragmentation of the GQR response.
Furthermore, by reducing smoothing parameter in the  calculations, we  
observe  the appearance of fine structure on top of the global fragmentation
in the strength functions.  A comparison with high resolution experiment shows 
that part of the observed peaks energies are located very close to the 
calculated energies. 

Our study demonstrates the importance of  coupling to low-lying surface modes for the 
understanding of fine structures in collective response. For this purpose,
the extended mean-field description  that includes both the incoherent and 
and the coherent mechanisms in an consistent manner appears as a promising
tool for the understanding of fine-structure properties in the fragmentation of giant
resonance excitations. It will be interesting to carry our similar investigations at finite
temperature.

{\bf Acknowledgments}
      
We thank A. Richter for providing the (e,e') data.                          
One of us (S. A.) gratefully acknowledges GANIL Laboratory for a partial
support and warm hospitality extended to him during his visits to Caen. This
work is supported in part by the US DOE grant No. DE-FG05-89ER40530.



\begin{table}
\begin{center}
\begin{tabular}{l|rrr|rr}
$^{40}$Ca / 0$^{+} $ & RPA & (c) & (i) & Theory & Experiment \\
                     &     &     &     &  (c+i) &      \\
\hline 
\hline 
$\overline{E_1}$-\scriptsize{[0-40]}    & 21.1 & 21.1 & 20.4 & 20.4 &   \\
$\overline{E_3}$-\scriptsize{[0-40]}    & 22.6 & 22.7 & 22.0 & 22.2 &   \\
$\overline{\Gamma}$-\scriptsize{[0-40]} & 4.6(4.0)  & 4.8(4.2)  & 4.7(4.1)   & 
4.9(4.4) &   \\
\hline
$\overline{E_1}$-\scriptsize{[8-29]}    & 20.6 & 20.5 & 19.8 & 19.8 & 18.9(0.1) 
($\alpha,\alpha')$ \protect{\cite{You97}} \\
$\overline{E_3}$-\scriptsize{[8-29]}    & 21.4 & 21.4 & 20.7 & 20.7 &  21.3(0.12) 
($\alpha,\alpha')$ \protect{\cite{You97}} \\
$\overline{\Gamma}$-\scriptsize{[8-29]} & 3.4(3.1) & 3.5(3.2)  & 3.4(3.1) & 
3.5(3.2) &  4.70(0.11) ($\alpha,\alpha')$ \protect{\cite{You97}} \\
\hline
$\%$ EWSR & $\%$ & $\%$ & $\%$ & $\%$ & $\%$ \\
\hline 
\scriptsize{$[0-40]$}       & 87.9 & 88.4 & 84.1 & 83.6  & \\
{\scriptsize $[12.5-22.5]$} & 54.2 & 54.4 & 63.8 & 56.8 & 50
($\alpha,\alpha')$ \protect{\cite{You97}} \\
\scriptsize{$[22.5-28.5]$} & 25.6 & 25.6 & 18.8 & 18.6  & 34.7 ($\alpha,\alpha')$
\protect{\cite{You97}}\\
\scriptsize{$[7.5-28.8]$}  & 80.6 & 80.9 & 77.1 & 76.2  & 92
($\alpha,\alpha')$ \protect{\cite{You97}} \\
\scriptsize{$[11-19]$}  & 21.6 & 22.1 & 26.9 & 27.8  & 44.2 $\pm$ 8.8
(e,e' $\alpha$)\protect{\cite{Koh98}} \\
\scriptsize{$[10.5-20]$}  & 29.1 & 30.0 & 36.4 & 37.0  & 30 $\pm$ 6
($\alpha,\alpha'$) \protect{\cite{Bra83}} \\
\end{tabular} 
\end{center}
\caption{Properties of GMR in $^{40}$Ca. 
Top: calculated mean-energy $\overline{E_1}$ and $\overline{E_3}$
and width $\overline{\Gamma}$ obtained by integrating moments of 
the strength in different energy intervals. Calculations are
carried out within RPA, and by including the coherent, the incoherent
and the coherent plus incoherent damping mechanisms, which are
indicated in columns under (c), (i) and (c+i), respectively. 
When available experimental data are also reported in the right 
column. We also display the calculated width for $\eta_s = 100 keV$
in parenthesis. Bottom: percentage of the EWSR calculated in 
different energy intervals. In the right column, if available, 
the corresponding experimental sum rules  are also reported 
together with the reactions and references.}
\label{tab0p:Ca40}
\end{table}

\begin{table}[tbp]
\begin{center}
\begin{tabular}{l|rrr|rr}
$^{40}$Ca / 1$^{-} $ & RPA & (c) & (i) & Theory & Experiment \\
                     &     &     &     &  (c+i) &      \\
\hline 
\hline 
$\overline{E_1}$-\scriptsize{[0-40]}     & 18.8 & 19.1 & 18.0 & 17.8 &   \\
$\overline{E_3}$-\scriptsize{[0-40]}     & 20.3 & 21.1 & 20.3 & 20.9 &   \\
$\overline{\Gamma}$-\scriptsize{[0-40]}  & 4.4(3.7) & 4.9(4.4) & 5.0(4.6) & 
5.5(5.0)   &   \\ 
\hline
E$_{peak}$ & 16.7 & 16.7 & 14.6 & 14.6 & \\
           & 18.6 & 18.4 & 16.4 & 17.2 & $\overline{E}=$19.0 \protect{\cite{Ahr72}} \\
  & & & & & $\Gamma=$ 4.0 \protect{\cite{Ahr72}} \\                     
\hline
$\%$ EWSR & $\%$ & $\%$ & $\%$ & $\%$ & $\%$ \\
\hline 
\scriptsize{$[0-40]$}     & 99.3 & 98.3 & 88.9 & 91.5  & \\
{\scriptsize $[10-21.5]$} & 71.0  & 71.2  & 65.7 & 65.1 & 58(15)
(e,e')\protect{\cite{Die94}} \\
&&&&& 63.0 ($\gamma,x$) \protect{\cite{Ahr75,Kam93}} \\
\scriptsize{$[21.5-40]$} & 27.9   & 26.8 & 22.6 & 25.9  & 
30.6 ($\gamma,x$) \protect{\cite{Ahr75,Kam93}}\\
\end{tabular} 
\end{center}
\caption{Same as table \protect{\ref{tab0p:Ca40}} for the GDR in 
$^{40}$Ca. In addition, in middle panel, $E_{peak}$ indicates the 
positions of the main peaks of the calculated strengths, and the
experimental peak position and the width of giant resonances 
are denoted by $\overline{E}$ and $\Gamma$.}
\label{tab1m:Ca40}
\end{table}

\begin{table}[tbp]
\begin{center}
\begin{tabular}{l|rrr|rr}
$^{40}$Ca / 2$^{+} $ & RPA & (c) & (i) & Theory & Experiment \\
                     &     &     &     &  (c+i) &      \\
\hline 
\hline 
$\overline{E_1}$-\scriptsize{[0-40]}    & 17.7 & 17.9 & 16.9  & 17.3 &   \\
$\overline{E_3}$-\scriptsize{[0-40]}    & 18.8 & 19.6 & 18.5  & 19.4 &   \\
$\overline{\Gamma}$-\scriptsize{[0-40]} & 3.5(2.6)  & 4.3(3.6) & 4.0(3.4) & 4.8(4.2) &   \\
\hline
$\overline{E_1}$-\scriptsize{[10-16]}   & 14.6 & 14.8 & 15.0 & 14.3 &   \\
$\overline{E_3}$-\scriptsize{[10-16]}   & 14.8 & 14.9 & 15.1 & 14.4 &   \\
\hline
\hline
$\overline{E_1}$-\scriptsize{[16-22]}    & 17.6 & 18.7 & 17.7 & 18.6 &   \\
$\overline{E_3}$-\scriptsize{[16-22]}    & 17.7 & 18.9 & 17.8 & 18.8 &   \\
\hline
E$_{peak}$ & 17.2 & 15.4 & 15.7 & 14.4 &  13.5 \protect{\cite{Zwa85,Die94}} \\
           &      & 17.6 &      & 17.6 &  18.0 \protect{\cite{Zwa85,Die94}} \\
           &      & 20.0 &      & 19.3 &                                   \\
\hline
$\%$ EWSR & $\%$ & $\%$ & $\%$ & $\%$ & $\%$ \\
\hline 
\scriptsize{$[0-40]$}   & 92.4 & 86.6 & 84.7 & 87.6 & \\
\scriptsize{$[13.2-15.2]$} & 23.4  & 12.7  & 10.7  & 11.1 & 
7.6 $\pm$ 1.1 (p,p') \protect{\cite{Yam87}}  \\
\scriptsize{$[13.2-16]$} & 29.1 & 16.3  & 13.3 & 14.3 & 
24.9 $\pm$ 5 (p,p') \protect{\cite{Lis89}}  \\
{\scriptsize $[10-16]$} & 0.0 & 26.6 & 39.1 & 31.0 & 
33 $\pm$ 7 $\left( e, e' x  \right)$ \protect{\cite{Die94,Die95}} \\
& & & & & 60(15) $\left( \alpha, \alpha ' \alpha_0  \right)$ \protect{\cite{Zwa85}}  \\
&&&&& (compilation from  \protect{\cite{Die95}})  \\
\scriptsize{$[16-22]$} & 72.5 & 51.6 & 34.7 & 33.6 & 
28.6 $\pm$ 7 (p,p') \protect{\cite{Ber79}}  \\
&  &  &  &  & $\left( \alpha, \alpha ' \alpha_0  \right)$ $\sim$ 40  \protect{\cite{Zwa85}} \\
&  &  &  &  &  44 (p,p') \protect{\cite{Lis89}} \\
\end{tabular} 
\end{center}
\caption{Same as table \protect{\ref{tab0p:Ca40}} for the GQR in $^{40}$Ca.}
\label{tab2p:Ca40}
\end{table}


\begin{table}[tbp]
\begin{center}
\begin{tabular}{l|rrr|rr}
$^{90}$Zr / 0$^{+} $ & RPA & (c) & (i) & Theory & Experiment \\
                     &     &     &     &  (c+i) &      \\
\hline 
\hline 
$\overline{E_1}$-\scriptsize{[0-40]}    & 19.0 & 19.2 & 18.8 & 19.0 &   \\
$\overline{E_3}$-\scriptsize{[0-40]}    & 20.0 & 20.5 & 20.0 & 20.4 &   \\
$\overline{\Gamma}$-\scriptsize{[0-40]} & 3.6(2.8)  & 4.0(3.3) & 3.9(3.1) &  
4.2(3.5) &   \\
\hline
E$_{peak}$ & 17.9 & 17.7 & 17.2 & 17.1 & 
$\overline{E} \simeq$ 16.0
\protect{\cite{Shl93}} \\
  & & & & & $\Gamma \simeq 3.3$ \protect{\cite{Shl93}} \\
\hline
$\%$ EWSR & $\%$ & $\%$ & $\%$ & $\%$ & $\%$ \\
\hline 
\scriptsize{$[0-40]$}   & 96.6 & 98.6 & 93.1 & 95.6 & \\
\scriptsize{$[12-20]$}  & 66.3 & 70.5 & 69.3 & 65.3 & 44 $\pm$ 20 
(p,p') \protect{\cite{Ber79}} \\
&&&&& 64 $\pm$ 14 $(\alpha,\alpha')$ \protect{\cite{You81}} \\
&&&&& 86 $\pm$ 15 ($^{17}$O+$^{90}$Zr) \protect{\cite{Lig93}} \\
&&&&& 83 $\pm$ 14  ($^{20}$Ne+$^{90}$Zr) \protect{\cite{Suo89}} \\
&&&&& 55 $\pm$ 13 ($^{40}$Ar+$^{90}$Zr) \protect{\cite{Suo90}} \\
\end{tabular} 
\end{center}
\caption{
Same as table \protect{\ref{tab0p:Ca40}} for the GMR in $^{90}$Zr.}
\label{tab0p:Zr90}
\end{table}

\begin{table}[tbp]
\begin{center}
\begin{tabular}{l|rrr|rr}
$^{90}$Zr / 1$^{-} $ & RPA & (c) & (i) & Theory & Experiment \\
                     &     &     &     &  (c+i) &      \\
\hline 
\hline 
$\overline{E_1}$-\scriptsize{[0-40]}    & 16.8 &17.2 &16.4 & 16.7 &   \\
$\overline{E_3}$-\scriptsize{[0-40]}    & 18.2 & 19.0 & 18.7 & 19.3  &   \\
$\overline{\Gamma}$-\scriptsize{[0-40]} & 3.8(3.0) & 4.4(3.8) & 4.8(4.2) 
& 5.2(4.7) &   \\
\hline
E$_{peak}$ & 15.7 &15.3 & 14.4 & 14.3 &  \\
           & 17.9 &17.8 &      &      &  $\overline{E}=$16.85 
\protect{\cite{Ber75}} \\
  & & & & & $\Gamma=$ 4.0 \protect{\cite{Ber75}} \\
\hline
$\%$ EWSR & $\%$ & $\%$ & $\%$ & $\%$ & $\%$ \\
\hline 
\scriptsize{$[0-40]$}  & 96.6 & 99.7 & 90.9 & 93.1 & \\
\scriptsize{$[11-19]$}  & 74.5  & 69.7 & 66.7 & 63.7 & 
 57 $(\gamma,x)$  \protect{\cite{Ber67}} \\
&&&&& 68 $(\gamma,x)$  \protect{\cite{Lep71}} \\
&&&&& 53 $\pm$ 13  $(\alpha,\alpha')$  \protect{\cite{Ber80}} \\
&&&&& 63 $\pm$ 14 ($^{20}$Ne+$^{90}$Zr)  \protect{\cite{Suo89}} \\
&&&&& 70 $\pm$ 28 ($^{40}$Ar+$^{90}$Zr)  \protect{\cite{Suo90}} \\
\end{tabular} 
\end{center}
\caption{
Same as table \protect{\ref{tab0p:Ca40}} for the GDR in $^{90}$Zr.}
\label{tab1m:Zr90}
\end{table}

\begin{table}[tbp]
\begin{center}
\begin{tabular}{l|rrr|rr}
$^{90}$Zr / 2$^{+} $ & RPA & (c) & (i) & Theory & Experiment \\
                     &     &     &     &  (c+i) &      \\
\hline 
\hline 
$\overline{E_1}$-\scriptsize{[11-25]}     & 16.8 & 16.8 & 16.2 & 16.4 &   \\
$\overline{E_3}$-\scriptsize{[11-25]}     & 17.0 & 17.5 & 16.7 & 17.2 &   \\
$\overline{\Gamma}$-\scriptsize{[11-25]}&1.9(1.4)&2.6(2.4)&2.3(1.9)&2.8(2.6)&   \\
\hline
\hline 
$\overline{E_1}$-\scriptsize{[0-8]}     & 5.2 & 5.2 & 3.9 & 3.9 &   \\
$\overline{E_3}$-\scriptsize{[0-8]}     & 5.6 & 5.4 & 4.5 & 4.6 &   \\
$\overline{\Gamma}$-\scriptsize{[0-8]}&1.3(0.6)&1.3(0.6)&1.3(0.7)&1.3(0.7)&   \\
\hline
E$_{peak}$ & 5.6  & 5.6  & 3.8  & 3.8  &  \\  
           & 9.5  & 9.5  & 9.5  & 9.5  &  \\  
           &      & 13.9 &      & 13.7 &  \\  
           & 16.4 & 16.2 & 15.5 & 15.6 &  $\overline{E}=$ 14.1(0.5) \protect{\cite{Ber80}} \\
  & & & & & $\Gamma=$ 4.0(0.5)  \protect{\cite{Ber80}} \\
\hline
$\%$ EWSR & $\%$ & $\%$ & $\%$ & $\%$ & $\%$ \\
\hline 
\scriptsize{$[0-40]$}      & 84.3 & 85.1 & 80.2 & 80.9 & \\
\scriptsize{$[11-25]$}     & 71.7 & 63.3 & 71.6 &  59.1 & \\
\scriptsize{$[0-8]$}       & 4.6  & 4.6  & 3.1  & 3.1   & \\
\scriptsize{$[10-18]$}     & 57.5 & 44.7  & 53.7  & 43.6  & 42 
$\pm$ 10 (p,p') \protect{\cite{Ber79}}\\
&&&&& 46 $\pm$ 9  $(\alpha,\alpha')$  \protect{\cite{Ber80}} \\
&&&&& 46 $\pm$ 14 ($^{20}$Ne+$^{90}$Zr)  \protect{\cite{Suo89}} \\
&&&&& 23 $\pm$ 14 ($^{40}$Ar+$^{90}$Zr)  \protect{\cite{Suo90}} \\
\end{tabular} 
\end{center}
\caption{
Same as table \protect{\ref{tab0p:Ca40}} for the GQR in $^{90}$Zr.}
\label{tab2p:Zr90}
\end{table}
 

\begin{table}[tbp]
\begin{center}
\begin{tabular}{l|rrr|rr}
$^{120}$Sn / 0$^{+} $ & RPA & (c) & (i) & Theory & Experiment \\
                     &     &     &     &  (c+i) &      \\
\hline 
\hline 
$\overline{E_1}$-\scriptsize{[0-40]}    & 17.4 & 17.8 & 17.1 & 17.6 &   \\
$\overline{E_3}$-\scriptsize{[0-40]}    & 18.6 & 19.7 & 18.6 & 19.7 &   \\
$\overline{\Gamma}$-\scriptsize{[0-40]} &3.6(2.7)&4.6(4.0)&3.9(3.1) &4.8(4.2)  &   \\
\hline
E$_{peak}$ & 16.8 & 16.4 & 16.1 & 15.9  &  \\  
& 18.5 & & 17.7 &  &  
$\overline{E} \simeq$15.3 
\protect{\cite{Shl93}} \\
  & & & & & $\Gamma \simeq$ 3.7
\protect{\cite{Shl93}} \\
\hline
$\%$ EWSR & $\%$ & $\%$ & $\%$ & $\%$ & $\%$ \\
\hline 
\scriptsize{$[0-40]$}  & 95.9 & 94.7 & 93.8  & 91.7 & \\
\scriptsize{$[8-20]$}  & 90.2 & 65.5 & 77.1  & 65.0 & \\
\scriptsize{$[12-20]$} & 78.7 & 63.6 & 75.0  & 63.0 & 
$ 61 \pm$ 15  (p,p') \protect{\cite{Ber79}}\\
&&&&& 72 $\pm$ 14  $(\alpha,\alpha')$  \protect{\cite{Ber80}} \\
&&&&& 120  $(\alpha,\alpha')$ \protect{\cite{You81}} \\
&&&&& 74  $\pm$ 15 $(\alpha,\alpha')$ \protect{\cite{Duh88}} \\
&&&&&  64.6 $\pm$ 14 $(\alpha,\alpha')$ \protect{\cite{Sha88}} \\
&&&&& 94. $\pm$ 20 ($^{17}$O+$^{120}$Sn) \protect{\cite{Lig93}} \\
\end{tabular} 
\end{center}
\caption{
Same as table \protect{\ref{tab0p:Ca40}} for the GMR in $^{120}$Sn.}
\label{tab0p:Sn120}
\end{table}

\begin{table}[tbp]
\begin{center}
\begin{tabular}{l|rrr|rr}
$^{120}$Sn / 1$^{-} $ & RPA & (c) & (i) & Theory & Experiment \\
                     &     &     &     &  (c+i) &      \\
\hline 
\hline 
$\overline{E_1}$-\scriptsize{[0-40]}     & 14.8 & 15.2 & 14.5 & 14.9 &   \\
$\overline{E_3}$-\scriptsize{[0-40]}     & 16.6 & 18.0 & 17.0 & 18.2 &   \\
$\overline{\Gamma}$-\scriptsize{[0-40]}  & 4.4(3.7)&5.3(4.8)& 4.8(4.3)&5.6(5.1)& \\
\hline
$\overline{E_1}$-\scriptsize{[0-8]}     & 3.7 & 3.7 & 3.7 & 3.7 &   \\
$\overline{E_3}$-\scriptsize{[0-8]}     & 5.2 & 5.3 & 5.3 & 5.3 &   \\
\hline
$\overline{E_1}$-\scriptsize{[10-25]}   & 15.2 & 15.2 & 14.8 & 14.9 &   \\
$\overline{E_3}$-\scriptsize{[10-25]}   & 15.9 & 16.0 & 15.6 & 15.8 &   \\
\hline
E$_{peak}$ & 2.7  & 2.7  & 2.7  & 2.7  &  \\
           & 10.3 & 10.3 & 10.3 & 10.2 &  \\
           & 14.3 & 13.4 & 13.6 & 13.2 &  \\
           & 16.9 & 18.0 &      & 17.9 & $\overline{E}=$15.4 \protect{\cite{Ber75}} \\
  & & & & & $\Gamma=$ 4.89  \protect{\cite{Ber75}} \\
\hline
$\%$ EWSR & $\%$ & $\%$ & $\%$ & $\%$ & $\%$ \\
\hline 
\scriptsize{$[0-40]$}      & 99.0 & 101.5 & 97.2 & 99.6 & \\
\scriptsize{$[0-8]$}       & 1.3  & 1.3 & 0.8 & 0.8  & \\
\scriptsize{$[8-20]$}      & 85.7 & 80.6 & 81.4 & 78.1 & \\
\scriptsize{$[13-18]$}     & 66.1 & 55.3 & 57.1 & 47.8 & 62 $(\gamma,x)$ \protect{\cite{Ful69}}\\
\end{tabular} 
\end{center}
\caption{
Same as table \protect{\ref{tab0p:Ca40}} for the GDR in $^{120}$Sn.}
\label{tab1m:Sn120}
\end{table}

\begin{table}[tbp]
\begin{center}
\begin{tabular}{l|rrr|rr}
$^{120}$Sn / 2$^{+} $ & RPA & (c) & (i) & Theory & Experiment \\
                     &     &     &     &  (c+i) &      \\
\hline 
\hline 
$\overline{E_1}$-\scriptsize{[0-40]}     & 15.8 & 16.3 &15.6  & 16.1  &   \\
$\overline{E_3}$-\scriptsize{[0-40]}     & 17.8 & 19.6 & 18.1 & 19.7 &   \\
$\overline{\Gamma}$-\scriptsize{[0-40]}  & 4.8(4.2) & 6.1(5.7) & 5.3(4.8) & 6.3(6.0) &   \\
\hline
$\overline{E_1}$-\scriptsize{[0-8]}      & 5.8 & 5.8 & 4.9 & 4.9 &   \\
$\overline{E_3}$-\scriptsize{[0-8]}      & 6.3 & 6.3 & 5.5 & 5.6 &   \\
\hline
$\overline{E_1}$-\scriptsize{[10-25]}    & 16.6 & 16.6 & 16.3 & 16.4 &   \\
$\overline{E_3}$-\scriptsize{[10-25]}    & 17.1 & 17.5 & 16.9 & 17.3 &   \\
\hline
E$_{peak}$ & 6.0  & 6.0  & 4.7  & 4.7  &  \\   
           &      & 12.8 &      & 12.6 & $\overline{E}=$13.3(0.3) \protect{\cite{Ber80}} \\
  & & & & & $\Gamma=$ 3.7(0.5) \protect{\cite{Ber80}}  \\   
           & 16.6 & 15.6 & 15.7 & 15.5 & \\
\hline
$\%$ EWSR & $\%$ & $\%$ & $\%$ & $\%$ & $\%$ \\
\hline 
\scriptsize{$[0-40]$}      & 69.8 & 68.1 & 68.0 & 67.0 & \\
\scriptsize{$[0-8]$}       & 2.7 & 2.7 & 2.2 & 2.3 & \\
\scriptsize{$[8-20]$}      & 55.1 & 43.3 & 52.5 & 43.0 & \\
\scriptsize{$[10-16]$} & 40.0 & 38.4  & 41.5  & 38.3 & 
53 $\pm$ 13  (p,p') \protect{\cite{Ber79}}\\
&&&&& 41 $\pm$ 9  $(\alpha,\alpha')$  \protect{\cite{Ber80}} \\
&&&&& 36 $\pm$ 6 
($^{17}$O+$^{120}$Sn) \protect{\cite{Lig93}} \\
\end{tabular} 
\end{center}
\caption{
Same as table \protect{\ref{tab0p:Ca40}} for the GQR in $^{120}$Sn.}
\label{tab2p:Sn120}
\end{table}


\begin{table}[tbp]
\begin{center}
\begin{tabular}{l|rrr|rr}
$^{208}$Pb / 0$^{+} $ & RPA & (c) & (i) & Theory & Experiment \\
                     &     &     &     &  (c+i) &      \\
\hline 
\hline 
$\overline{E_1}$-\scriptsize{[0-40]}    & 15.3 & 16.1 & 15.0 & 15.7 &   \\
$\overline{E_3}$-\scriptsize{[0-40]}    & 16.5 & 18.5 & 16.8 & 18.6 &   \\
$\overline{\Gamma}$-\scriptsize{[0-40]} & 3.4(2.5)  & 4.9(4.4) & 4.0(3.3) & 5.3(4.7)
 &    \\
\hline 
$\overline{E_1}$-\scriptsize{[0-20]}    & 14.6 & 14.4 & 14.0 & 13.9 &   \\
$\overline{E_3}$-\scriptsize{[0-20]}    & 15.0 & 14.9 & 14.5 & 14.5  &   \\
$\overline{\Gamma}$-\scriptsize{[0-20]} & 2.2(1.4)  & 2.4(1.8) & 2.2(1.5) & 2.4(1.9)
& \\
\hline
E$_{peak}$ & 16.3 & 14.2 & 13.9 & 13.6 & 
$\overline{E} \simeq$13.6
\protect{\cite{Van91,Shl93}} \\
  & & & & & $\Gamma \simeq$ 2.5
\protect{\cite{Van91,Shl93}} \\
\hline
$\%$ EWSR & $\%$ & $\%$ & $\%$ & $\%$ & $\%$ \\
\hline 
\scriptsize{$[0-40]$}      & 99.2 & 102.6 & 96.8 & 100.1 & \\
\scriptsize{$[0-20]$}      & 88.5 & 70.7& 75.9 & 65.8 & \\
\scriptsize{$[10-16]$}     & 66.0 & 58.4  & 68.8  & 59.0 & 
63 $\pm$ 17  (p,p') \protect{\cite{Ber79}}\\
&&&&& 64 $\pm$ 13  $(\alpha,\alpha')$  \protect{\cite{Ber80}} \\
&&&&& 61 $\pm$ 13  $(\alpha,\alpha')$  \protect{\cite{You81}} \\
&&&&& 59    $(\alpha,\alpha')$  \protect{\cite{Mor80}} \\
&&&&& 69  (p,p')  \protect{\cite{Ber86}} \\
&&&&& 64   ($^3$He,$^3$He)  \protect{\cite{Bue80}} \\
&&&&& 84 $\pm$ 30 ($^{17}$O+$^{208}$Pb) \protect{\cite{Bee90}} \\
&&&&& 92 $\pm$ 8 ($^{17}$O+$^{208}$Pb) \protect{\cite{Lig93}} \\
\end{tabular} 
\end{center}
\caption{Same as table \protect{\ref{tab0p:Ca40}} for the GMR in $^{208}$Pb.}
\label{tab0p:Pb208}
\end{table}

\begin{table}[tbp]
\begin{center}
\begin{tabular}{l|rrr|rr}
$^{208}$Pb / 1$^{-} $ & RPA & (c) & (i) & Theory & Experiment \\
                     &     &     &     &  (c+i) &      \\
\hline 
\hline 
$\overline{E_1}$-\scriptsize{[0-40]}    & 13.2 & 13.7 & 12.8 & 13.3 &   \\
$\overline{E_3}$-\scriptsize{[0-40]}    & 15.3 & 16.8 &15.9 & 17.0 &   \\
$\overline{\Gamma}$-\scriptsize{[0-40]} & 4.2(3.5)&5.2(4.7)&4.9(4.3) & 5.6(5.1) &   \\
\hline
E$_{peak}$ & 2.1  & 2.1  & 2.1  & 2.1  &  \\
           & 13.0 & 12.5 & 11.6 & 11.4 &  \\
           & 16.7 & 16.7 & 16.7 & 16.7 &  $\overline{E}=$ 13.5 \protect{\cite{Ber75}} \\
  & & & & & $\Gamma=$ 4.0  \protect{\cite{Ber75}} \\
\hline
$\%$ EWSR & $\%$ & $\%$ & $\%$ & $\%$ & $\%$ \\
\hline 
\scriptsize{$[0-40]$}       & 99.2 & 101.8 & 95.9  & 97.2 & \\
\scriptsize{$[0-5]$}        & 0.1  & 0.1 & 0.1 & 0.1 & \\
\scriptsize{$[5-15]$}       & 70.7 & 59.8 & 65.0 & 59.2 & \\
\scriptsize{$[15-25]$}      & 23.9 & 32.3 & 23.2 & 28.0 & \\
\scriptsize{$[10-17]$}      & 75.5 & 66.0 & 66.1 & 61.0 & 
68  $(\gamma,x)$ \protect{\cite{Har64}} \\
&&&&& 91 $(\gamma,x)$ \protect{\cite{Vey70}} \\
&&&&& 90 $(\gamma,x)$ \protect{\cite{You72}} \\
&&&&& 68  (p,p')  \protect{\cite{Ber86}} \\
\end{tabular} 
\end{center}
\caption{Same as table \protect{\ref{tab0p:Ca40}} for the GDR in $^{208}$Pb.}
\label{tab1m:Pb208}
\end{table}

\begin{table}[tbp]
\begin{center}
\begin{tabular}{l|rrr|rr}
$^{208}$Pb / 2$^{+} $ & RPA & (c) & (i) & Theory & Experiment \\
                     &     &     &     &  (c+i) &      \\
\hline 
\hline 
$\overline{E_1}$-\scriptsize{[0-40]}     & 11.5 & 12.4 & 11.2 & 12.0 &   \\
$\overline{E_3}$-\scriptsize{[0-40]}     & 14.6 & 17.6 & 15.4 & 18.0 &   \\
$\overline{\Gamma}$-\scriptsize{[0-40]}  &4.9(4.4)&6.6(6.4)&5.6(5.2)&7.0(6.8) &   \\
\hline 
$\overline{E_1}$-\scriptsize{[8-20]}     & 13.2 & 13.0 & 12.5 & 12.6  &   \\
$\overline{E_3}$-\scriptsize{[8-20]}     & 13.5 & 13.9 & 13.0 & 13.6 &   \\
$\overline{\Gamma}$-\scriptsize{[8-20]}  &1.7(1.1)&2.8(2.6)&1.9(1.5)&2.8(2.7) &   \\
\hline 
$\overline{E_1}$-\scriptsize{[0-8]}      & 5.1 & 5.1 & 4.4& 4.4 &   \\
$\overline{E_3}$-\scriptsize{[0-8]}      & 5.9 & 5.9 & 5.1 & 5.1 &   \\
$\overline{\Gamma}$-\scriptsize{[0-8]}  &1.8(1.6)&1.8(1.6)&1.6(1.2)&1.6(1.2) &   \\
\hline 
E$_{peak}$  & 6.2  & 6.1  & 4.9  &  4.8 &  \\    
            & 13.0 & 11.2 & 12.1 & 10.8 &  $\overline{E}=$10.9 \protect{\cite{Ber80}} \\
  & & & & & $\Gamma=$  2.4(0.4) \protect{\cite{Ber80}} \\   
            &      & 13.8 &      & 13.8 &  \\    
\hline
$\%$ EWSR & $\%$ & $\%$ & $\%$ & $\%$ & $\%$ \\
\hline 
\scriptsize{$[0-40]$}    & 92.4 & 93.4   & 89.0 & 90.6  & \\
\scriptsize{$[0-8]$}     & 10.7  & 10.5  & 9.2  & 9.1  & \\
\scriptsize{$[8-12.5]$}  & 15.5  & 25.2  & 39.0 & 27.5  &
49  $\pm$ 12      (p,p') \protect{\cite{Ber79}}\\
&&&&& 44 $\pm$ 10 (p,p') \protect{\cite{Dan86}} \\
&&&&& 44 $\pm$ 8  (p,p') \protect{\cite{Ada86}}   \\
&&&&& 62 $\pm$ 11 (p,p') \protect{\cite{Ber86}}   \\
&&&&& 50 $\pm$ 8  (O+Pb,H-I) \protect{\cite{Bee90}}   \\
&&&&& 50 $\pm$ 5 $ (\alpha,\alpha')$  \protect{\cite{Ber80}} \\
&&&&&   33 $\pm$ 8 ($^{17}$O+$^{208}$Pb) \protect{\cite{Lig93}} \\
&&&&& 36 $\pm$ 9 ($^{20}$Ne+$^{208}$Pb) \protect{\cite{Suo89}} \\
&&&&& 36 $\pm$ 12  ($^{40}$Ar+$^{208}$Pb) \protect{\cite{Suo90}} \\
\end{tabular} 
\end{center}
\caption{Same as table \protect{\ref{tab0p:Ca40}} for the GQR in $^{208}$Pb.}
\label{tab2p:Pb208}
\end{table}

\begin{table}
\begin{center}
\begin{tabular}{rrrrrr}
(e,e')  \protect{\cite{Kuh81}}     & (p,p') \protect{\cite{Ber86}} & 
(p,p')  \protect{\cite{Kam97-2}}   & (c+i) & (c) & (i) \\
\hline
8.9   & 8.9     & 8.9  &                        &         &    \\
9.4   & 9.3     & 9.4  & 9.3  & 9.3  & 9.3  \\
9.6   &         & 9.6  & 9.9  & 9.9  &      \\
10.1  &         & 10.1 & 10.2 & 10.3 &      \\
10.7  & 10.6    & 10.7 & 10.7 & 10.8 &      \\
11.5  &         & 11.0 & 11.0 & 11.3 &      \\
      &         &      &      &      & 11.9 \\
\end{tabular}  
\end{center}
\caption{Experimental energies of peaks position observed in the 
GQR of $^{208}$Pb in (p,p') and (e,e') experiments. Calculated
peak positions including the coherent, the incoherent and
the coherent plus incoherent self-energies are reported under columns
indicated by (c), (i) and (c+i), respectively.} 
\label{tab:fine}
\end{table}


\begin{references}

\bibitem{Van91} {A. van der Woude, {\it Electric 
and Magnetic Giant Resonances in
Nuclei} (World Scientific, 1991) p. 101 (and reference therein), 
edited by J. Speth.}

\bibitem{Kuh81} {G. K\"uhner, D. Meuer, S. M\"uller, A. Richter,
E. Spamer, O. Titze and W. Kn\"upfer, Phys. Lett. {\bf 104B} (1981) 189.}

\bibitem{Kam97-2} {S. Kamerdzhiev, J. Lisantti, P. von Neumann-Cosel, A.
Richter, G. Tertychny and J. Wambach, Phys. Rev. {\bf C55} (1997) 2101.}

\bibitem{Lac00-2} {D. Lacroix, A. Mai, P. von Neumann-Cosel, A. Richter and J.
Wambach, Phys. Lett. {\bf B479} (2000) 15.}


\bibitem{Ber83}  {G. F. Bertsch, P.F. Bortignon and R. A. Broglia, Rev. Mod.
Phys. {\bf 55} (1983) 287.}

\bibitem{Ber94}  {G.F. Bertsch and R.A. Broglia, {\it Oscillation in finite
quantum systems}, Cambridge (1994).}

\bibitem{Dro90}  {S. Drozdz, S. Nishizaki, J. Speth and J. Wambach,
Phys. Rep. {\bf 197} (1990) 1.}

\bibitem{Lac99} {D. Lacroix and Ph. Chomaz, Phys. Rev. {\bf C60} (1999) 064307.}


\bibitem{Goe82}  {K.Goeke and P.-G.Reinhard, {\it ''Time-Dependent
Hartree-Fock and Beyond''}, Proceedings, Bad Honnef, Germany (1982).}


\bibitem{Abe95}  {Y. Abe, S. Ayik, P. G. Reinhard and E. Suraud, Phys. Rep. 
{\bf 275} (1996) 49.}


\bibitem{Bor81} {P.F. Bortignon and R. Bloglia, Nucl. Phys. 
{\bf A371} (1981) 405. }

\bibitem{Ayi98}  {S. Ayik, O. Yilmaz, A. Gokalp and P. Schuck, Phys. Rev. 
{\bf C58} (1998), 1594.}

\bibitem{Yil99} {O. Yilmaz, A. Gokalp, 
S. Yildirim and S. Ayik, Phys. Lett. {\bf B472} (2000) 258. }   

\bibitem{Bor86}  {P. Bortignon, R. A. Broglia, G. F. Bertsch and J. Pacheco, Nucl. Phys.
{\bf A460} (1986) 149.}

\bibitem{Gio98}  {N. Giovanardi, P.F. Bortignon and R.A. Broglia, Nucl. Phys.
{\bf A641} (1998) 95.}

\bibitem{Kam95} {S. Kamerdzhiev, J. Speth and G. Tertychny, 
Phys. Rev. lett. {\bf 72} (1995) 1995.}

\bibitem{Kam97-1} {S. Kamerdzhiev, J. Speth and G. Tertychny, 
Nucl. Phys. {\bf A624} (1997) 328.}

\bibitem {Deb92}  F.V. De Blasio, W. Cassing, M. Tohyama, P.F.
Bortignon, and R. Broglia, Phys. Rev. Lett. \textbf{68} (1992) 1663.

\bibitem{Lac98}  {D. Lacroix, Ph. Chomaz and S. Ayik, Phys. Rev. {\bf C 58}
(1998) 2154.}
\bibitem{Ayi00-1} {S. Ayik, 
D. Lacroix and Ph. Chomaz, Phys. Rev. C {\bf 61} (2000) 014608.}

\bibitem{Cho00}  {Ph. Chomaz, D. Lacroix, S. Ayik and M. Colonna, Phys.
Rev. {\bf C62} (2000) 024307.}

\bibitem{Lac00}  {D. Lacroix, Ph. Chomaz and S. Ayik, Phys.
Lett. {\bf B489} (2000) 137.}

\bibitem{Ayi00}  { S. Ayik, preprint YITP-00-23 and  
Phys. Lett. {\bf B}  (2000) {\it in press}.}

\bibitem{AyiAbe}
S.  Ayik and  Y. Abe, preprint YITP-00-55 and 
submitted to Phys. Rev. {\bf C}  (2000).

\bibitem{Ayi94}
S.  Ayik, Z. Phys. {\bf A350} (1994) 45.

\bibitem{Ivanov}
Y. B. Ivanov and S. Ayik, Nucl. Phys. {\bf A593} (1995) 233.


\bibitem{Gua96}  { A. Guarnera, M. Colonna and P. Chomaz, Phys. Lett 
\bf{B373}, (1996) 267.}

\bibitem{Fra00}  {J. D. Frankland {\it et al}, Preprint nucl-ex/0007019,
Preprint nucl-ex/0007020.}



\bibitem{Rin80}  {P. Ring and P. Schuck, {\it The nuclear many-body problem,
Spring-Verlag,} New-York (1980).}



\bibitem{Van81}  {N. Van Giai and H. Sagawa, Nucl. Phys. {\bf A371} (1981) 1.}

\bibitem{Cha98}  E. Chabanat, P. Bonche, P. Haensel, J. Meyer and R.
Schaeffer, Nucl. Phys. {\bf A627} (1997) 710; Nucl. Phys. {\bf A635} (1997)
231.{\ Erratum: Nucl. Phys. {\bf A643} (1998) 441.}

\bibitem{Sag84}  {H. Sagawa and G.F. Bertsch, Phys. Lett. {\bf B146} (1984)
138.}                                        

\bibitem{Bar85} {M. Barranco, A. Polls, S. Marcos, J. Navarro and J. Treiner,
 Phys. Lett. {\bf B154} (1985) 96.}                                        

\bibitem{Sat87} {G.R. Satchler, Nucl. Phys. {\bf A472} (1987) 215. }

\bibitem{Hor95} { D.J. Horen, J. R. Beene and G. R. Satchler, 
Phys. Rev. {\bf C52} (1995) 1554 (and reference therein).}

\bibitem{You97} {D. H. YoungBlood, Y.-W. Lui and H. L. Clark, 
Phys. Rev. {\bf C55} (1997) 2811. }

\bibitem{Bra83} {S. Brandenburg {\it et al}, Phys. Lett. {\bf B130} (1983) 
9. S. Brandenburg {\it et al}, Nucl. Phys. {\bf A466} 
(1987) 29.}

\bibitem{Koh98} {M Kohl, P. von Neumann-Cosel, A. Richter, G. Schrieder and S.
Strauch, Phys. Rev. {\bf C57} (1998) 3167. }

\bibitem{Ahr72} {J. Hahrens {\it et al }, 
Proc. of the Int. Conf. on nuclear structure, Sendai, Japan, 1972.}

\bibitem{Die94} {H. Diesener {\it et al}, Phys. Rev. Lett. {\bf 72} (1994) 
1994.}

\bibitem{Ahr75} {J. Ahrens {\it et al}, Nucl. Phys. {\bf A251} (1975) 479.}

\bibitem{Kam93} {S. Kamerdzhiev, J. Speth, G. Tertychny and V. Tselyaev,
Nucl. Phys. {\bf A555} (1993) 90. (Table 1).}

\bibitem{Zwa85} {F. Zwarst, A.G. Drentje, M.N. Harakeh and A. Van der Woude,
Nucl. Phys. {\bf A439} 117.}

\bibitem{Yam87} {T. Yamagata {\it et al}, Phys. Rev. {\bf C36} (1987) 
573.}
\bibitem{Lis89} {J. Lisantti {\it et al}, Phys. Rev. {\bf C40} (1989) 
211.}

\bibitem{Die95} {H. Diesener {\it et al}, Phys. Lett. {\bf B352} (1995) 
201.}

\bibitem{Ber79} {F.E. Bertrand, G.R. Satchler, D.J. Horen
and A. van der Woude, Phys. Lett. {\bf B198} (1979) 198.}

\bibitem{Shl93} { S. Shlomo and D.H. Youngblood, Phys. Rev. {\bf C47} 
(1993) 529 (and reference therein).}

\bibitem{You81} { D.H. Youngblood, P. Kogucki, J.D. Bronson, U. Garg, 
Y.-W. Lui and C. M. Rozsa, Phys. Rev. {\bf C23} (1981) 1997.}

\bibitem{Lig93} {R. Liguori Neto {\it et al}, Nucl Phys. {\bf A560} (1993) 
733.}

\bibitem{Suo89} {T. Suomijarvi {\it et al}, Nucl Phys. {\bf A491} (1989)
314.}
\bibitem{Suo90} {T. Suomijarvi {\it et al}, Nucl Phys. {\bf A509} (1990)
369.}

\bibitem{Ber75} {B.L. Berman and S. C. Fultz, Rev.  Mod. Phys. {\bf 47} (1975) 713. }

\bibitem{Ber67} {B. L. Berman, J. T. Cadwell, R. R. Harvey, M.A. Kelly,
R. L. Bramblett and S. C. Fultz, Phys. Rev. {\bf 162} (1967) 1098.}  

\bibitem{Lep71} {A. Lepr\^etre, H. Beil, R. Berg\`ere, P. Carlos, A. Veyssi\`ere
and M. Sugarawa, Nucl. Phys {\bf A175} (1971) 609.}

\bibitem{Ber80} {F.E. Bertrand et al, Phys. Rev. {\bf C22} (1980) 1832. }

\bibitem{Duh88} {G. Duhamel, M. Buenerd, P. de Saintignon, J. Chauvin,
D. Lebrun, Ph. Martin and G. Perrin, Phys. Rev. {\bf C38} (1988) 2509.}

\bibitem{Sha88} {M. M. Sharma, W. T. A. Borghols, S. Brandenburg, S. Crona, A.
Van der Woude and M. H. Harakeh, Phys. Rev. {\bf C38} (1988) 2562.}
                
\bibitem{Ful69} {S. C. Fultz, B. L. Berman, J. T. Caldwell, R. L. Bramlett and
M. A. Kelly, Phys. Rev. {\bf 186} (1969) 1255.}

\bibitem{Mor80} {H. P. Morsch, C. \"uk\"osd, M. Rogge, P. Turek, H. Machner and
C. Mayer-B\"oricke, Phys. Rev. {\bf C22} (1980) 489.
H. P. Morsch, M. Rogge, P. Turek, C. Mayer-B\"oricke and P. Decowski
Phys. Rev. {\bf C25} (1982) 939. }

\bibitem{Ber86} {F.E. Bertrand, {\it et al}, Phys. Rev. {\bf C34} (1986) 
45.}                                  

\bibitem{Bue80} {M. Buenerd, D. Lebrun, Ph. Martin,
P. de Saintignon and G. Perrin, Phys. Rev. Lett. {\bf 45} (1980) 1670.}

\bibitem{Bee90} {J.R. Beene {\it et al}, Phys. Rev. {\bf C41} (1990) 
920.}                                  

\bibitem{Har64} {R. R. Harvey, J. T. Caldwell, R. L. Bramlett and 
S. C. Fultz, Phys. Rev. {\bf 136} (1964) B126.}

\bibitem{Vey70} {A. Veyssi\`ere, H. Beil, R. Berg\`ere, P. Carlos, 
A. Lepr\^etre, Nucl. Phys. {\bf A159} (1970) 561.}

\bibitem{You72} {L. M. Young, PhD Thesis (1972), 
University of Illinois (unpublished).}
\bibitem{Ber75} {B.L. Berman and S. C. Fultz, Rev. Mod. Phys. {\bf 47} (1975) 
713.}

\bibitem{Dan86} {D.K McDaniels {\it et al}, Phys. Rev. {\bf C33} (1986) 
1943.}
\bibitem{Ada86} {G.S. Adams {\it et al}, Phys. Rev. {\bf C33} (1986) 
2054.}                                  

\bibitem{Lis91} {J. Lisantti, E. J. Stephenson, A. D. Bacher, 
P. Li, R. Sawafta, P. Schwandt, S. P. Wells, S. W. Wissink, W. Unkelbach, and
J. Wambach, Phys. Rev. C {\bf 44}, R1233 (1991).}

\bibitem{Dro00} {A. Drouart, PhD Thesis SPhN-Saclay, France, {\it to be
published}.}
\end{references}
\end{document}